\documentclass[prd,a4paper,eqsecnum,nofootinbib,preprint,floatfix]{revtex4}
\usepackage{colordvi}
\usepackage{graphicx}
\usepackage{bm}
\usepackage{multirow}
\usepackage{enumerate}
\reversemarginpar
\def\<{\langle}
\def\>{\rangle}
\newcommand{\text}{\rm}

\def\Eq#1{Eq.~(\ref{#1})}

\newcommand\alb{{a\sqrt{\lambda/(2\pi)}}}

\def\Eq#1{Eq.~(\ref{#1})}

\begin{document}

\vspace*{0.7in}
 
\begin{center}
{\large\bf Solving two-dimensional large-$N$ QCD\\
with a nonzero density of baryons and arbitrary quark mass.
}

\vspace*{1.0in}
\vspace{-1cm}
{Barak Bringoltz\\
\vspace*{.2in}
Department of Physics, University of Washington, Seattle,
WA 98195-1560, USA\\
}
\end{center}

\begin{abstract}

We solve two-dimensional large-$N$ QCD in the presence of a nonzero baryon number $B$, and for arbitrary quark mass $m$ and volume $L$. We fully treat the dynamics of the gluonic zero modes and check how this affects results from previous studies of the $B=0$ and $B=1$ systems. For a finite density of baryons, and for any $m>0$, we find that the ground state contains a baryon crystal with expectation values for $\<\bar \psi \gamma_\mu \psi\>$ that have a helix-like spatial structure. We study how these evolve with $B$ and see that the volume integral of $\<\bar \psi \psi\>$ strongly changes with the baryon density. We compare this emerging crystal structure with the sine-Gordon crystal, which is expected to be a good approximation for light quarks, and find that it is a very good approximation for surprisingly heavy quarks. We also calculate the way the ground state energy $E$ changes as a function of the baryon number $B$, and find that for sufficiently large densities the function $E(B)$ is well described by the equation of state for free massless quarks, thus suggesting a quark-Hadron continuity. From $\partial E(B)/\partial B$ we calculate the quark chemical potential $\mu$ as a function of $B$ and see that the baryons repel each other. The way $\mu$ depends on $B$ also allows us to translate our findings to the grand-canonical ensemble. The resulting phase structure along the $\mu$-axis contains a phase transition that occurs at a value of $\mu$ equal to the baryon mass divided in $N$, and that separates a $\mu$-independent phase with intact translation symmetry from a $\mu$-dependent phase with spontaneously broken translation symmetry. Finally, our calculations confirm the presence of a partial large-$N$ Eguchi-Kawai volume independence, as described in {\tt Phys.Rev.D79:105021}, that arises only if one treats the gluonic zero modes correctly. 

\end{abstract}


\maketitle

\setcounter{page}{1}
\newpage
\pagestyle{plain}

\section{Introduction}
\label{intro}

A first principles calculation of QCD properties in dense environments is a challenging problem facing the community working on non-perturbative aspects of QCD. This is mostly due to the severe sign problem inflicting lattice simulations at these densities. Trying to avoid the sign problem typically leads one to study relatively low densities \cite{Stephanov}. Other approaches such as analytic weak/strong coupling expansions come with their own shortcomings : weak coupling expansions \cite{cmQCD} rely on the smallness of the effective interaction between he quasi-particles of dense QCD, but these interactions really become weak only for asymptotically large chemical potentials \cite{SR}. On the other hand, lattice strong-coupling methods \cite{strong} are, of course, far from the continuum limit. Furthermore, the approach of using the gauge/gravity duality is also not free of difficulties \cite{ADS} : it is truly systematically controlled only when the curvature scale associated with the classical gravity solution is small in its natural units. Unfortunately, the classical solution for a single or few baryons has strong curvatures and can get $O(1)$ corrections from higher derivative terms in the effective low energy action, which are currently unknown. For a large number of baryons one also needs to assume small curvatures, and this restricts one to a translational invariant states which may have little to do with the true ground state.

Given the above theoretical status, we find it surprising that the soluble system of large-$N$ QCD in $1+1$ dimensions -- the `t Hooft model  \cite{thooftmodel} -- has not been generally solved for such densities. For example, the interesting works of Refs.~\cite{SchonThies} either restrict to zero quark mass, or expand in the value of the $m$. Unfortunately, for such values of $m$ the two-dimensional baryon is almost massless and may behave  very differently from a four-dimensional baryon. A different theoretical approach to this problem is presented in Ref.~\cite{GHN} and invokes the `Eguchi-Kawai' equivalence~\cite{EK}. We defer a short discussion on Ref.~\cite{GHN} to our summary.

The aim of this paper is therefore to solve two-dimensional large-$N$ QCD for general values of the quark mass $m$, the baryon number $B$, and the spatial volume $L$. 

The outline of this paper is as follows. Section~\ref{Hc} describes the results of our previous publication \cite{thooftaxial} where we derived an IR and UV regularized form of the `t Hooft classical Hamiltonian, and its corresponding `t Hooft equations. Besides depending on $m$, on $B$, and on $L$, this construction depends on the lattice spacing $a$ and on an additional technical parameter which we denote by $M$ and that needs to be taken to $\infty$. The role of $M$ is to discretize a certain continuous variable on which the solution of the `t Hooft equations depends. Comparing the construction of Ref.~\cite{thooftaxial} to other works such as Ref.~\cite{Salcedo} we see that these instead have set $M=1$. Such choice is inconsistent with the dynamics of the spatial glue fields, especially when the volume is small (see \cite{SchonThies_decompact}, \cite{LNT}, and \cite{thooftaxial}). Thus, in Section~\ref{psiBpsimB} we check what is the effect of increasing $M$ from $1$ on the properties of the ground state for $B=0$ and $B=1$, which were already calculated previously in Ref.~\cite{Salcedo}. Section~\ref{n_B} contains our main results and describes the way a crystal is formed if the system contains a variety of baryon numbers in the range $0\le B\le 30$. In Section~\ref{partial} we confirm the presence of a partial volume reduction anticipated in Ref.~\cite{thooftaxial}. Section~\ref{summary} summarizes our results.

\section{The classical `t Hooft Hamiltonian ${\cal H}_F$ and its minimization}
\label{Hc}

In Ref.~\cite{thooftaxial} we made a first step towards solving large-$N$ QCD in $1+1$ dimensions for arbitrary values of baryon number, volume, and quark mass. For the purpose of self-completeness we use the current section to repeat the main results of that paper. 

We derived a regularized form of the `t Hooft classical Hamiltonian which is both UV and IR finite by using the canonical formalism defined on a one-dimensional spatial lattice of spacing $a$ and volume $L=aL_s$ (here $L_s$ denotes the number of lattice sites). The fermion discretization of our choice was a single species of staggered fermions, which gives rise to a single Dirac fermion in the continuum (i.e. there is no residual doubling).\footnote{This regularization is self-consistent only if $L_s$ is chosen to be even.}

Since our system is IR regularized by a finite box, we cannot gauge away the spatial gluons completely. Specifically, a set of $N-1$ zero modes which correspond to the eigenvalues of the spatial `Polyakov loop' remain dynamical. Taking this into account and using the coherent state formalism of Ref.~\cite{YaffeCoherent} we obtained a form for the `t Hooft Hamiltonian that depends on the expectation values of the Polyakov loops and of the fermion bilinear operators. Minimizing this classical Hamiltonian in a $1/N$ expansion we saw that the Polyakov loops have zero expectation values and that the classical Hamiltonian becomes a functional of an infinite set of basis functions. Each such basis contains $L_s$ orthogonal scalar functions that span the $L_s-$dimensional space furnished by the spatial coordinate $x\in [1,L_s]$. We denote these functions by $\phi^n_a(x)$:
\begin{equation}
\phi^n_a(x) \qquad a=1,2,\dots,\infty \quad ; \quad n,x\in [1,L_s],
\end{equation}
with
\begin{eqnarray}
\sum_x \left(\phi^n_a(x)\right)^\star \, \phi^m_a(x) &=& \delta_{nm},\label{ortho}\\
 \sum_x \phi^n_a(x) \, \left(\phi^n_a(y)\right)^\star &=& \delta_{xy}. \label{ortho1}
\end{eqnarray}
In terms of the functions $\phi$ the regularized `t Hooft Hamiltonian becomes (here we also regularize the infinite number of functions to be the finite number $M$)
\begin{eqnarray}
{\cal H}_F(\phi)/N &=&  \lim_{M\to \infty} \left[ \frac1{M} \sum_a\sum_{x\in Z_{L_s}}\left\{  \left(-\frac{i}2 \rho^a_{x,x+1}  + c.c. \right)+ \hat m (-1)^x \rho^a_{xx}\right\} \right. \nonumber \\
&-& \frac{\hat \lambda}{4}  \frac1{L_sM^2} \sum'_{abl} \sum_{xy\in Z_{L_s}} \frac{\rho^a_{xy}\, \rho^b_{yx} \, e^{-i(x-y)\left(\frac{2\pi}{L_s}(\frac{a-b}{M}+l)\right)}}{4\sin^2\left(\frac{2\pi}{L_s}(\frac{a-b}{M}+l)/2\right)}\nonumber \\ \nonumber \\
&+& \left.\frac{\hat \lambda(B+L_s/2)}{4}  \frac1{L_sM^2} \sum'_{abl} \frac1{4\sin^2\left(\frac{2\pi}{L_s}(\frac{a-b}{M}+l)/2\right)}\right].\label{HKC4}
\end{eqnarray}
Here and below the bare mass $\hat m$ and bare `t Hooft coupling $\hat \lambda$ are dimensionless and related to the corresponding dimensional quantities $m,\lambda$ via
\begin{eqnarray}
\hat \lambda &=& a^2 \lambda,\\
\hat m &=& am.
\end{eqnarray}
Also, the set of $M$ `density matrices' $\rho^a_{xy}$ is given by
\begin{equation}
\rho^a_{xy}\equiv\sum_{n=1}^{B+L_s/2}\, \phi^n_a(x)\, \phi^{n\star}_a(y).\label{rho}
\end{equation}
The $B$-dependence of \Eq{rho}, together with the way ${\cal H}_F(\phi)$ depends on $B$, is the way the baryon number enters the discussion (in the Hamiltonian approach one constraints the quantum Hilbert space to obey Gauss law and for the $SU(N)$ gauge theory that we study this includes fixing the global $U(1)$ baryon charge $B$).\footnote{In our numerical studies we find that, while irrelevant for the minimization of ${\cal H}_F$, the last term in \Eq{HKC4} is crucial to include in order to get the correct baryon mass.} Finally, note that the prime on the sums means that the terms with $a=b$ and $l=L_s$ are excluded ($l$ generally gets integer values between $1$ and $L_s$). This exclusion is a direct result of our IR regularization and of the correct treatment of the zero modes. It is exactly this exclusion that becomes the principle value prescription often used in differently regularized treatments of the `t Hooft model.

Let us make an important remark already alluded to in the previous section: in \cite{thooftaxial} we showed that the choice of $M=1$ (instead of $M=\infty$), that was made elsewhere (for example in \cite{Salcedo}), is effectively equivalent to incorrectly treating the spatial gluonic zero modes and that it amounts to choosing the expectation values of the spatial Polyakov loops to be nonzero. This is wrong and leads to erroneous results. In particular, it implies that the volume dependence of the Hamiltonian is in contradiction with general arguments of the sort of \cite{EK} (for a detailed discussion on this point see \cite{thooftaxial}). In the current paper we test how increasing $M$ from $1$ to larger values of $O(15-25)$ fixes this problem, and we do so for different values of $B$. This extends the $B=0$ discussion of \cite{SchonThies_decompact} and corrects the results of Ref.~\cite{Salcedo} for $B=1$. 

 Going back to the derivation of the `t Hooft equations, we perform the variation of ${\cal H}_F$ with respect to the $M$ functions $\phi^n_a(x)$ 
\begin{equation}
\frac{\delta}{\delta (\phi^n_a(x))^\star} \left( {\cal H} - \sum_{m,b} \epsilon^b_m \sum_x \phi^m_b(x) \, \phi^{m\star}_b(x) \right)=0, \label{variation}
\end{equation}
and we find the they obey the following $M$ coupled nonlinear differential equations (here we used the Lagrange multiplier $\epsilon^a_{n}$ to enforce \Eq{ortho})
\begin{eqnarray}
\sum_{y\in Z_{L_s}} \, h^a_{xy} \ \phi^n_a(y) &=& \epsilon^a_n \, \phi^n_a(x),\label{diagEQ}
\end{eqnarray}
with
\begin{eqnarray}
h^a_{xy}&=& +\frac{i}2 \left(\delta_{y,x+1} - \delta_{y,x-1}\right) + \hat m\, (-1)^x\, \delta_{xy} -\hat \lambda \, v^a_{xy}, \label{h}\\
v^a_{xy}&=&\frac1{2M}\sum_b \, K_{ab}(y,x)\, \left(\sum_{m=1}^{B+L_s/2}\, \phi^m_b(x)\, \phi^{m\star}_b(y)\right),\\
K_{ab}(y,x) &=&  \frac1{L_s}\,\sum_{l\in Z_{L_s}}' \, \frac{e^{\frac{2\pi i(x-y)}{L_s}\left(\frac{a-b}{M}+l\right)}}{4\sin^2\left(\frac12\left(\frac{2\pi (a-b)}{ML_s} + \frac{2\pi l}{L_s}\right)\right)}.\label{Kernel}
\end{eqnarray}

Since $K_{ab}$ explicitly depends on $\phi$, solving \Eq{diagEQ} is a self-consistent process which we perform by beginning with a random choice for $\rho^a_{xy}$, inserting it into $h^a_{xy}$, diagonalizing the latter $M$ matrices to obtain the $M$ sets of eigenfunctions $\phi^n_a(x)$, and these then provide us with a new estimate of $\rho^a_{xy}$. This process is iterated until convergence is observed. The only restriction our initial random choice of $\rho$ needs to obey is 
\begin{equation}
\sum_x \rho^a_{xx} = B + L_s/2,
\end{equation}
which reflects \Eq{ortho1} and \Eq{rho}.
Within the space of all functions that obey \Eq{diagEQ} the correct solution is the one that has the lowest value of ${\cal H}_F$. The latter {\em is not} equal to $\sum_{an}\epsilon^a_n$, since this will count the Coulomb interaction twice. Instead one has
\begin{eqnarray}
{\cal H}_{\rm solution}/N &=& \frac1{2M}\sum_{n=1}^{B+L_s/2}\,\sum_{a=1}^M\left(\epsilon^n_a +  \sum_x \left[{\rm Im}\,  \left(\phi^n_a(x) \phi^{n\star}_a(x+1)\right) + \hat m\, (-1)^x\ \phi^n_a(x) \phi^{n\star}_a(x)\right] \right)\nonumber\\
&+&\frac{\hat \lambda(B+L_s/2)}{4}  \frac1{L_sM^2} \sum'_{abl} \frac1{4\sin^2\left(\frac{2\pi}{L_s}(\frac{a-b}{M}+l)/2\right)}.\label{Hsol}
\end{eqnarray}
Finally, to calculate the values of quark bilinears we build a single Dirac fermion from the staggered fields and what we get is
\begin{eqnarray}
\<\psi^\dag \psi\>(X) &=&\frac1{2M} \sum_{a=1}^{M} \,\,\left(\rho^a_{x,x} +  \rho^a_{x-1,x-1}\right), \label{lat2cont1}\\
\<\bar \psi \psi\>(X) &=& \frac1{2M} \sum_{a=1}^{M} \,\,\left(\rho^a_{x,x} -  \rho^a_{x-1,x-1}\right),\label{lat2cont2}\\
\<\bar \psi \,i\gamma_5 \psi\>(X) &=& -\frac1{M} \, \sum_{a=1}^{M} {\rm Imag}\left(\rho^a_{x,x+1} \right),\label{lat2cont3}\\
\<\bar \psi \gamma_1 \psi\>(X) &=& \frac1{M} \sum_{a=1}^{M}  {\rm Real}\left(\rho^a_{x,x+1}\right),\label{lat2cont4}
\end{eqnarray}
where here $X=ax/2, \,x={\rm even}$. In some of the plots that appear in forthcoming sections we present UV-regularized results for $\<\psi^\dag\gamma_\mu\psi\>(X)$. The regularization we choose is to subtract the corresponding values in the free theory. Specifically, anticipating large-$N$ volume independence, we subtract the free theory expectation values at infinite volume. In Appendix~\ref{appfree} we show that this means we should perform the following subtractions
\begin{eqnarray}
\<\psi^\dag \psi\>(X) &\stackrel{\rm reg.}{\longrightarrow}&\<\psi^\dag \psi\>(X) - \frac12  ,\label{nbfree}\\
\<\bar \psi \left(1,i\gamma_5\right) \psi\>(X) &\stackrel{\rm reg.}{\longrightarrow}& \<\bar \psi \left(1,i\gamma_5\right) \psi\>(X) - \<\bar \psi \left(1,i\gamma_5\right) \psi\>_{\rm free},\\
\<\bar \psi \psi\>_{\rm free} &=& 
-m\int_{p_F}^\pi \,\frac{dp}{2\pi}\,\frac1{\sqrt{\hat m^2 + \sin^2 (p/2)}},\label{condfree}\\
\<\bar \psi \, i\gamma_5\,\psi\>_{\rm free} &=& 
+\int_{p_F}^\pi \,\frac{dp}{2\pi}\,\frac{\sin^2(p/2)}{\sqrt{\hat m^2 + \sin^2 (p/2)}},\label{cond5free}
\end{eqnarray}
with $p_F = \pi \frac{B}{L_s/2}$. Also, we always find that $\<\bar \psi \gamma_1\psi\>=0$ which means that time-reversal symmetry is intact. This, of course, also happens in the free theory (see the appendix).\footnote{The way we choose to build the continuum quantities is slightly different from that of Ref.~\cite{Salcedo}. The difference, however, vanishes in the continuum limit.}

In the following sections we implement the numerical solution of the self-consistent Eqs.~(\ref{h})--(\ref{Kernel}) for different values of $B$, $m/\surd\lambda$, $L\surd\lambda$, and $a\surd \lambda$.

\section{The case of $B=0$ and $B=1$.}
\label{psiBpsimB}

\begin{figure}[p]
\centerline{
\includegraphics[width=12cm]{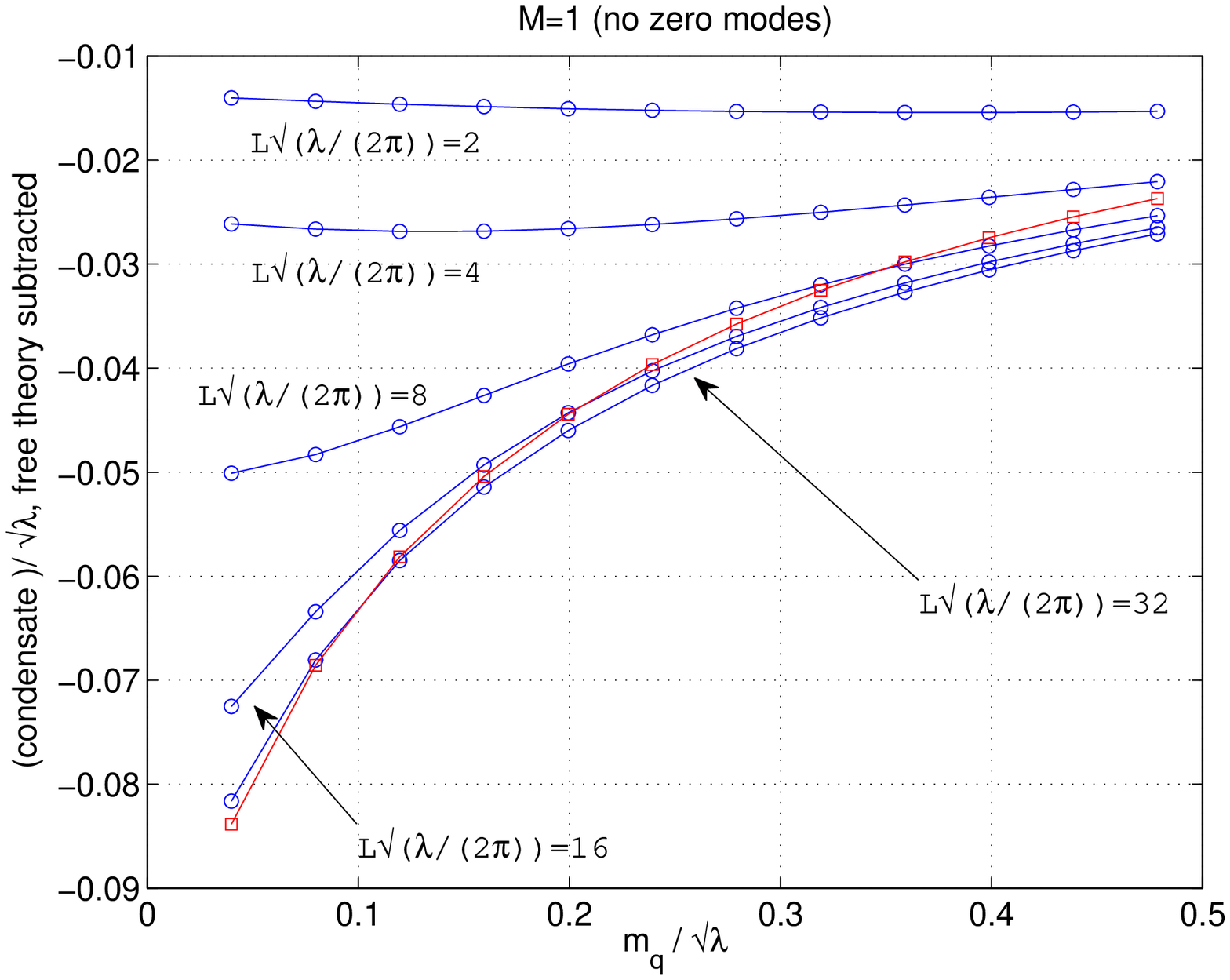}
}
\vskip 0.5cm 
\centerline{
\includegraphics[width=12cm]{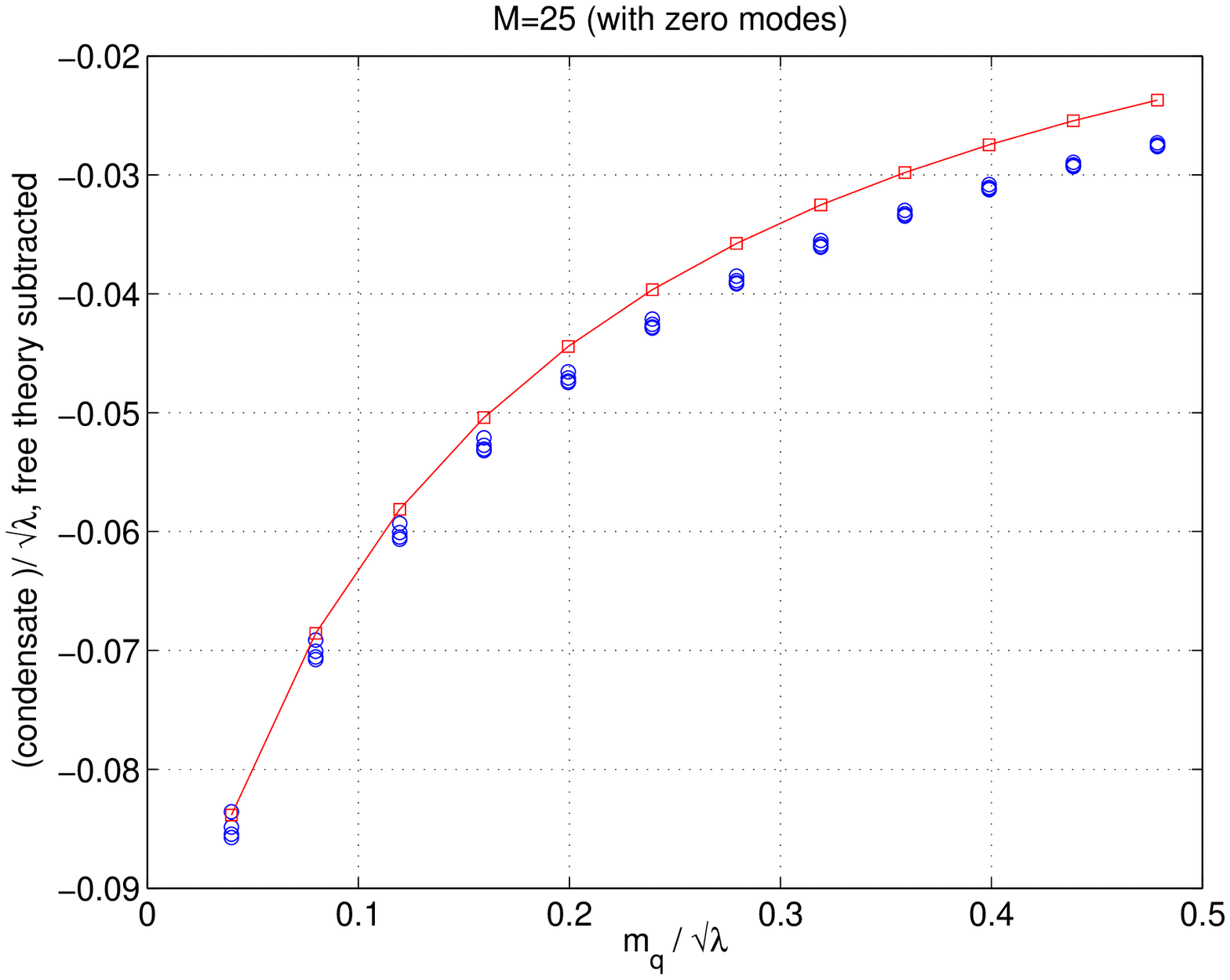}
}
\caption{Quark mass dependence of the condensate for $B=0$ and different volumes. \underline{Upper panel:} Mistreating the spatial gluonic zero modes (setting $M=1$). \underline{Lower panel:} Incorporating the zero modes with $M=25$. The red curve is the result of Ref.~\cite{Burkardt} (see text). As explained in the text, the discrepancy between our data in the left panel and the analytic curve reflects lattice artifacts.}
\label{pbp}
\end{figure}

In this section we wish to see how sensitive are the values of the quark condensate and baryon mass to the technical parameter $M$. In \cite{thooftaxial} we showed that the erroneous $M=1$ choice corresponds to setting the spatial Polyakov loops to unity while $M=\infty$ sets them to zero. Thus, we expect that the volume dependence in the former case to be strong, while in the latter it should disappear (for the connection between the expectation values of the Polyakov loops and volume dependence see \cite{EK,thooftaxial}.
 This expectation is confirmed by the results presented in Fig.~\ref{pbp}: both the plots on the upper and lower panels of the figure were obtained for $B=0$ and a relatively fine lattice spacing of $\alb =0.123$. The plots present the way the UV-regularized value of $\<\bar \psi\psi\> / N$ depends on the quark mass for different values of $L$.
The only parameter by which these plots  differ is $M$: it is fixed to $1$ on the top panel and to $25$ on the bottom. Fortunately, for the $B=0$ system there is an analytic prediction for arbitrary mass \cite{Burkardt} which we represent by the solid (red) curve in the figures and compare to our plots.
As anticipated, while the $L$-dependence of the upper plot is very strong, it is hardly visible on the lower panel. 
This is large-$N$ Eguchi-Kawai independence at work.

\begin{figure}[p]
\centerline{
\includegraphics[width=20cm]{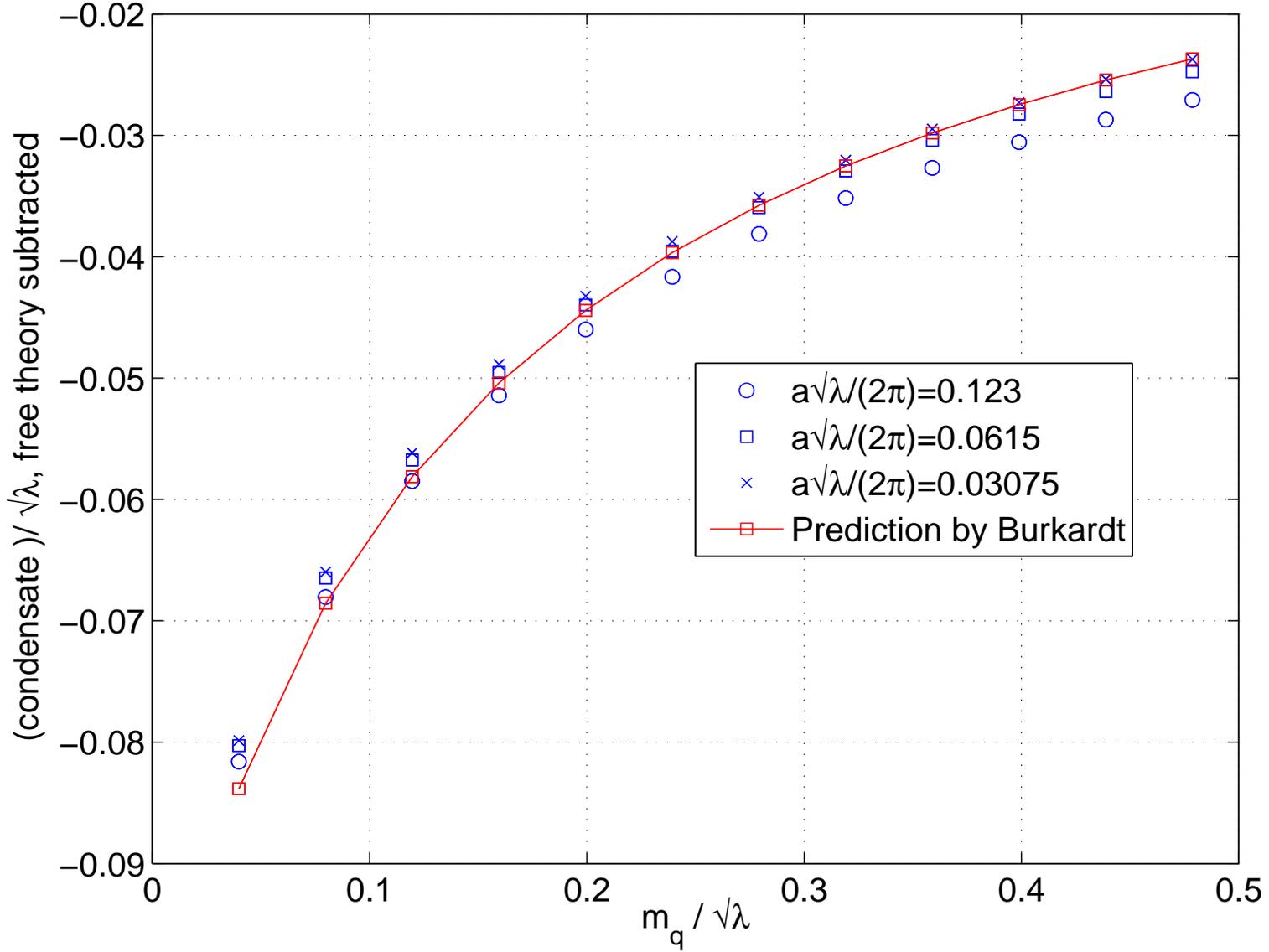}
}
\caption{Quark mass dependence of the condensate for $B=0$ and different lattice spacings. We use $M=1$, but a relatively large volume of $L\sqrt{\lambda/(2\pi)}=32$, where the zero modes do not induce any significant $1/M$ effect. The red curve is the exact result in the continuum and infinite volume limit from \cite{Burkardt}, which agrees very well with the results from the smallest lattice spacing of $a\sqrt{\lambda/(2\pi)}=0.03075$.}
\label{pbp_a}
\end{figure}

\begin{figure}[p]
\centerline{
\includegraphics[width=20cm]{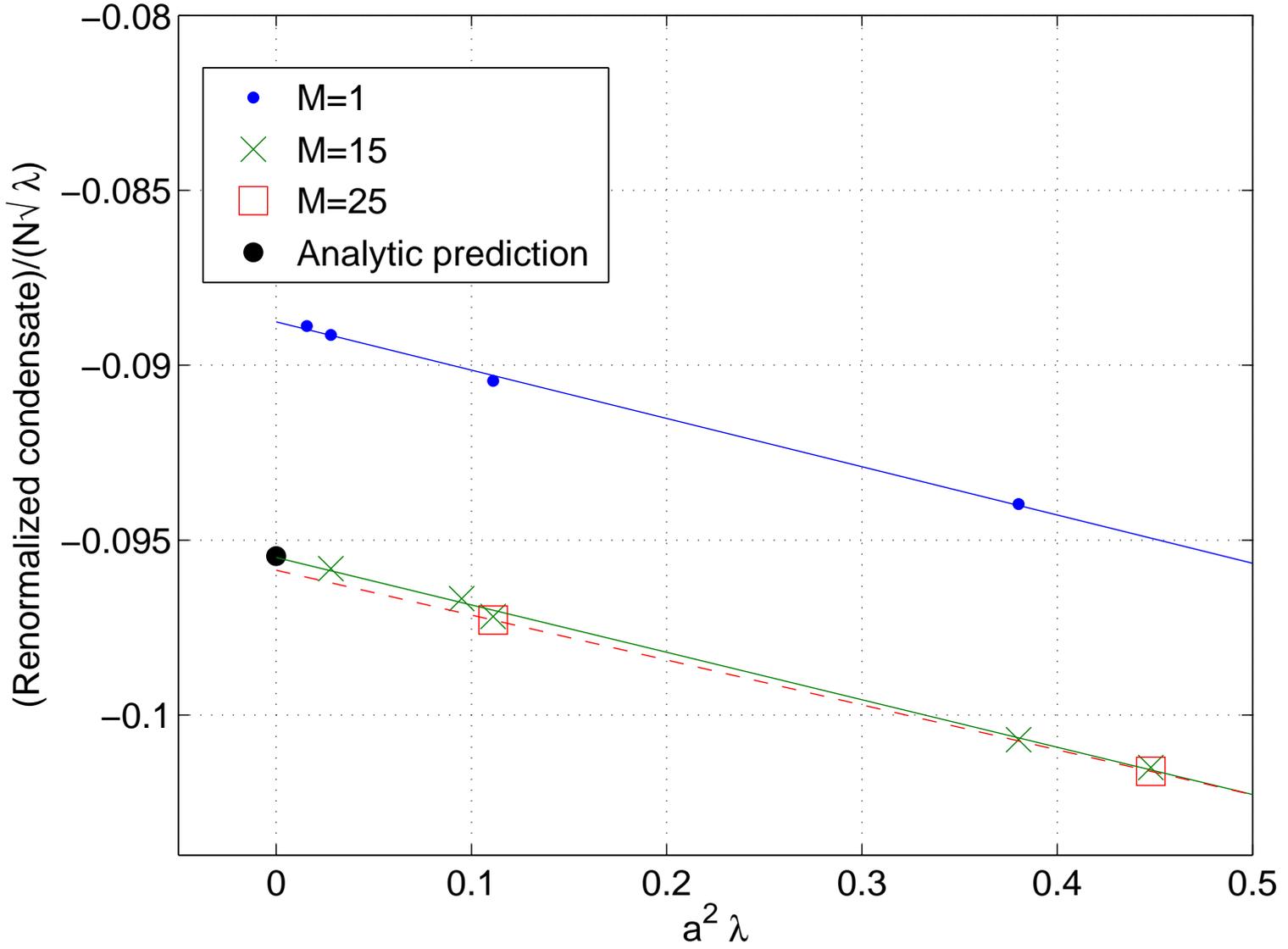}
}
\caption{Continuum extrapolation of the quark condensate for $m/\sqrt{\lambda}=0.05/\sqrt{2\pi}$ and for several values of the parameter $M$. The Filled circle (black) at $a=0$ is the analytic result \cite{Burkardt}. All data was obtained for the same volume of $L\sqrt{\lambda/(2\pi)}=32$.}
\label{cont_extr}
\end{figure}

A glance at the lower panel of Fig.~\ref{pbp} might be alarming: our data seems to be quite far from the analytic solution in the regime of large quark masses. This, however, is a lattice artifact as we demonstrate in Fig.~\ref{pbp_a}. There, we plot the $m$-dependence of the quark condensate for different lattice spacings, and for $L\sqrt{\lambda/(2\pi)}=32$ ($M$ was set to unity here since the $1/M$ corrections should be small for this volume). Indeed, we find that as $a\sqrt{\lambda/(2\pi)}$ drops from $0.123$ to $0.0615$ to $0.03075$, the deviations of the data from the prediction of \cite{Burkardt} decrease.
 We take the deviations of the $a\sqrt{\lambda/(2\pi)}=0.03075$ data from the analytic prediction to reflect the numerical convergence error of our calculation, the $1/M$, and $1/L$ corrections, as well as the numerical error involved in evaluating the analytic expressions of Ref.~\cite{Burkardt}. To check that these deviations are under control we plot in Fig.~\ref{cont_extr} the continuum extrapolation of $\<\bar \psi \psi\>/(N\surd\lambda)$ for very light quarks with $m/\sqrt{\lambda}=0.05/\sqrt{2\pi}$, a relatively large volume of $L\sqrt{\lambda/(2\pi)}=32$, and for several values of $M$. As the plot shows, while the data obtained for $M=15,25$ extrapolates to the analytic result in the continuum, the data of $M=1$ does not. This discrepancy of $\sim O(8\%)$ is thus a measure of the $1/M$ and $1/L$ corrections.

\begin{figure}[p]
\centerline{
\includegraphics[width=20cm]{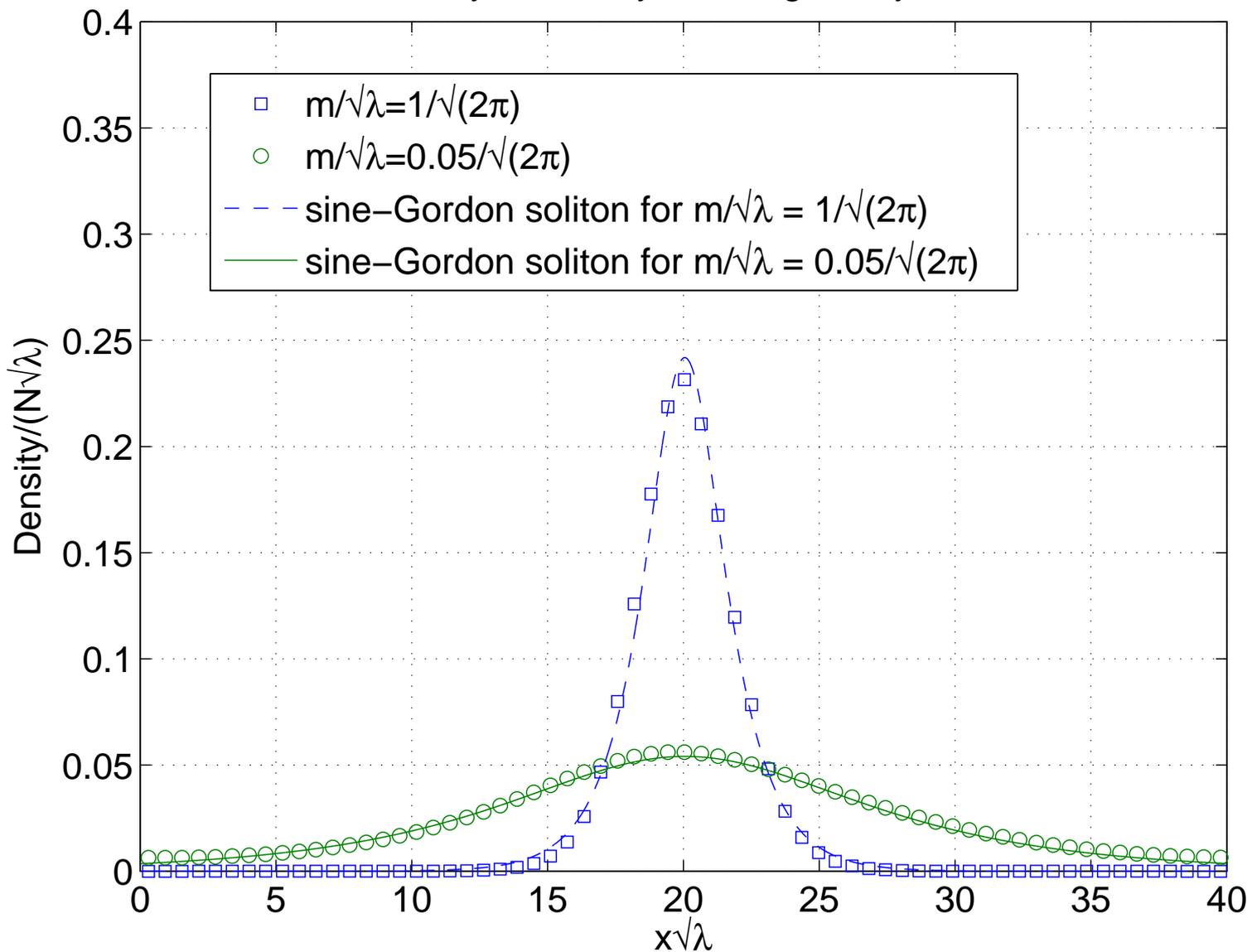}
}
\caption{Baryon density of a single baryon. \underline{Squares (blue):}  an intermediate quark mass; $m/\sqrt{\lambda}=1/\sqrt{2\pi}$. \underline{Circle (green):} Small quark mass; $m/\sqrt{\lambda}=0.05/\sqrt{2\pi}$. The dashed (blue) and solid (green) curves are the predictions of the sine-Gordon soliton that are expected to be descriptive of the data in the $m/\surd\lambda\to 0 $ limit.}
\label{nB1}
\end{figure}

Next we wish to examine the single baryon ground state. We begin by plotting, in Fig.~\ref{nB1}, the baryon density $\<\psi^\dag \psi\>/N$ as a function of the spatial coordinate $x$.\footnote{Here, in contrast to \Eq{HKC4}, we denote by $x$ the dimensional spatial coordinate (and not the lattice site index), i.e. from here on we denote the $X$ of Eqs.~(\ref{lat2cont1})--(\ref{lat2cont4}) by $x$.} We choose a moderately large volume of $L\sqrt{\lambda/(2\pi)}=16$, a lattice spacing of $L\sqrt{\lambda/(2\pi)}=0.123$, and present results for $m/\sqrt{\lambda/(2\pi)} = 0.05$ and $1$. These parameters are the same as those used in Ref.~\cite{Salcedo}. Our calculation differs from the one in that paper by our choice of $M$ : we use $M=25$ while Ref.~\cite{Salcedo} used $M=1$.
As seen in the Figure, the baryon becomes less localized as the quark mass drops. Eventually, at $m=0$, it will spread throughout space, and become a massless delocalized objects. We see that the other quark bilinears, $\<\bar\psi \psi\>$ and $\<\bar\psi i\gamma_5\psi\>$, also modulate in space, but we postpone discussing this modulation to the next section.

\begin{figure}[p]
\centerline{
\includegraphics[width=20cm]{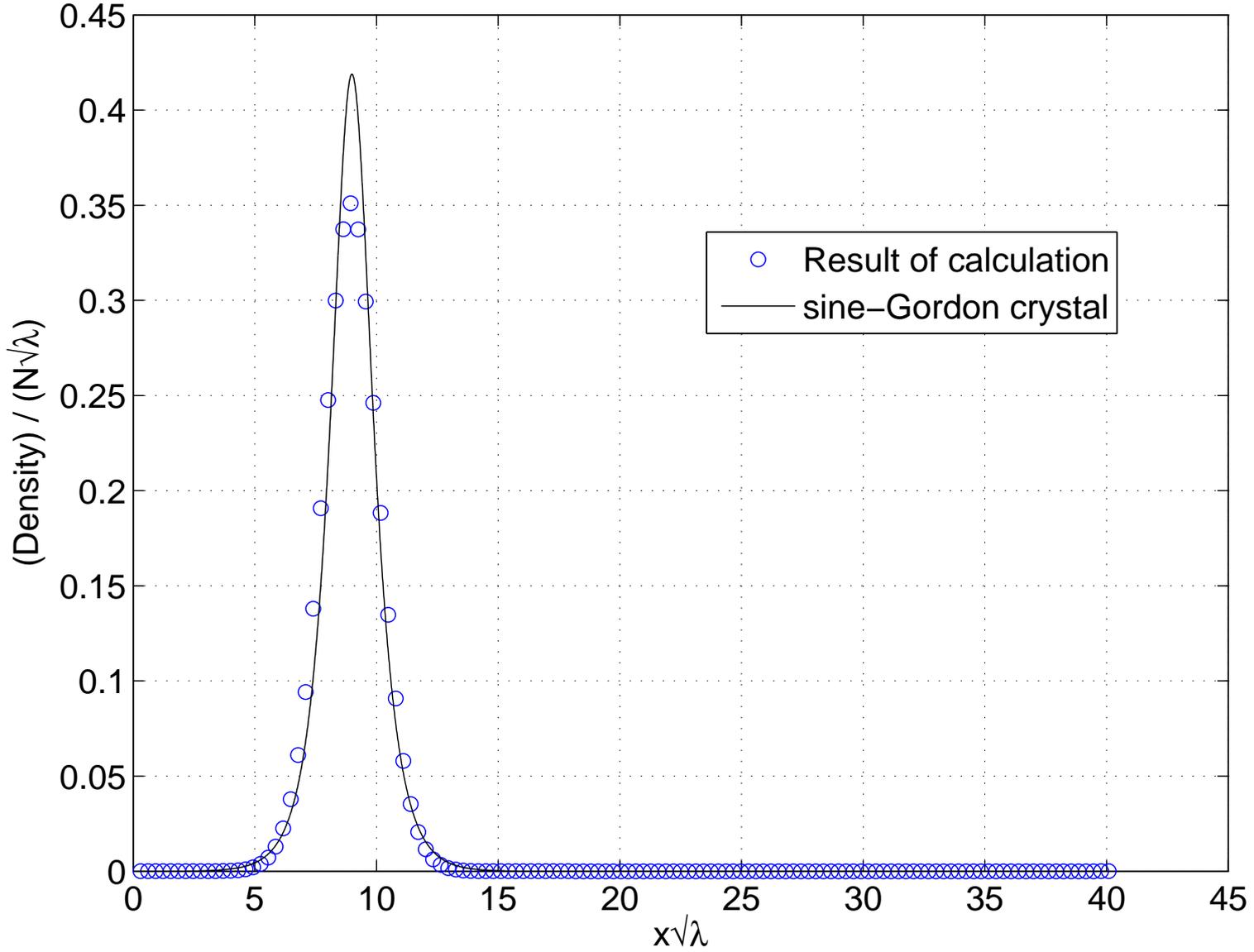}
}
\caption{Baryon density of a single baryon obtained for very heavy quarks :  $m/\sqrt{\lambda}=3/\sqrt{2\pi}$. Again the curve is the prediction of the sine-Gordon soliton valid in the light quarks regime of $m/\surd\lambda \ll 1$.}
\label{ultra_heavy}
\end{figure}

Also plotted in Fig.~\ref{nB1} is the prediction for the soliton density of a single soliton in the sine-Gordon model. As argued in \cite{Salcedo,SchonThies} this prediction should agree with the `t Hooft model's baryon density, but only for sufficiently small values of $m/\surd\lambda$. To emphasize the importance of having $m/\sqrt{\lambda}\ll 1$ for this argument we mention that, in the approximation made in Ref.~\cite{Salcedo,SchonThies}, the width of this soliton is determined by the value of $\<\bar \psi\psi\>$ in the chiral limit, and one neglects all finite-$m$ contributions to the latter. Remarkably, however, we see that the agreement between the prediction and our data is good even for $m/\sqrt{\lambda} \simeq 0.4$, where the value of $\<\bar \psi\psi\>$ differs by a factor of $4$ from its value in the chiral limit. This agreement is not expected. For even smaller quark masses of $m/\surd\lambda \simeq 0.02-0.1$ it was also a surprise to the authors of Ref.~\cite{Salcedo}. Here, and throughout the current paper, we see that such agreement seems good even for much heavier quarks than these. To make this point stronger we plot in Fig.~\ref{ultra_heavy} the single baryon density for very heavy quarks of $m/\surd\lambda \simeq 1.2$, where deviations from the soliton density are already  visible, but still very modest. Further surprising results, for even heavier quarks are presented in Section~\ref{n_B} where we discuss finite density.

\begin{figure}[p]
\centerline{
\includegraphics[width=20cm]{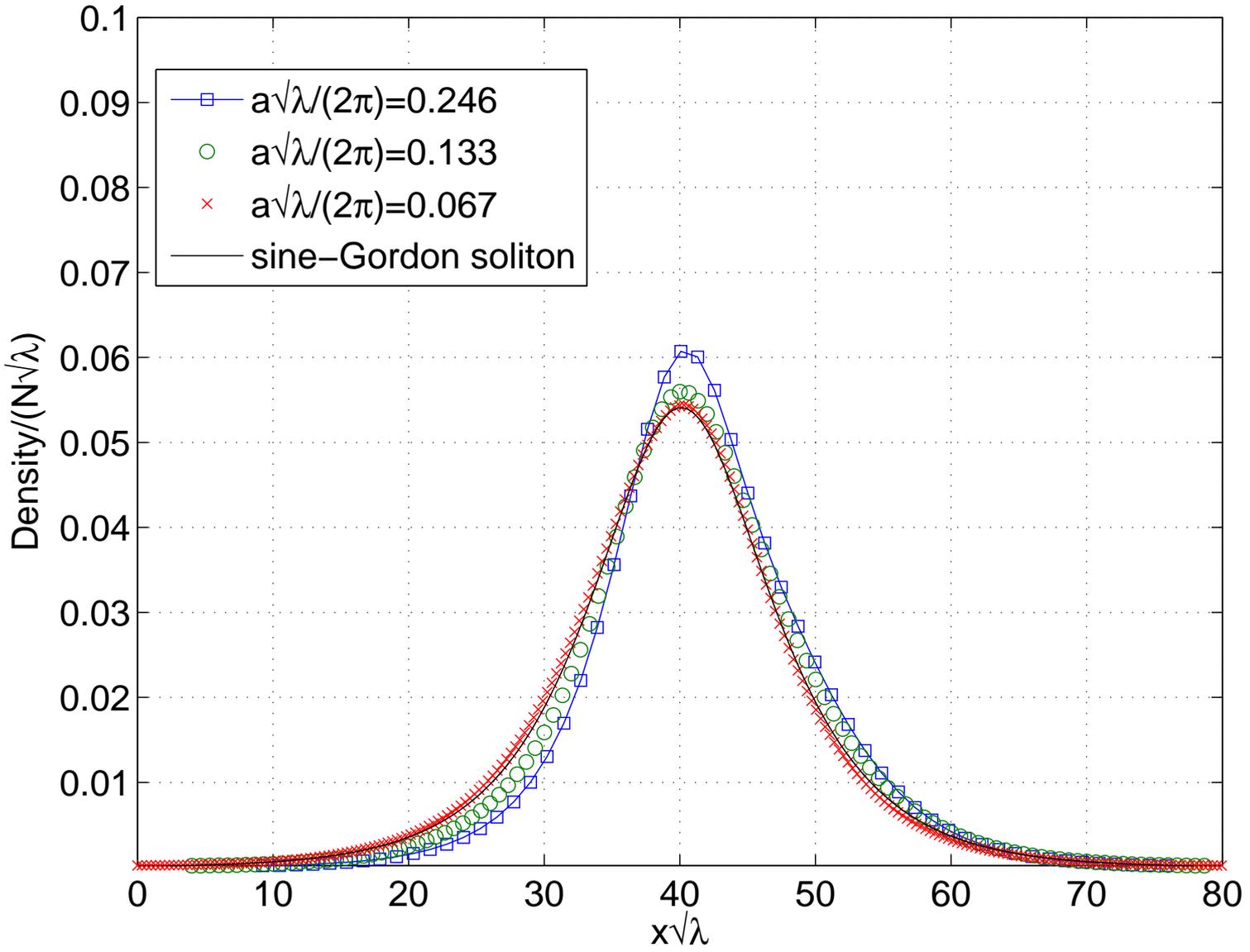}
}
\caption{Baryon density of a single baryon for different lattice spacings. The solid curve (black) is the result of the sine-Gordon. Note the different scales of the $x$ and $y$ axes compared to those of Fig.~\ref{nB1}.}
\label{nB_compare_a}
\end{figure}

To analyze the lattice corrections in our calculation we plot, in Fig.~\ref{nB_compare_a}, the baryon density of $B=1$, $m/\sqrt{\lambda} = 0.05/\sqrt{2\pi}$, $M=15$, $L\sqrt{\lambda/(2\pi)}=32$, and for three different lattice spacings. The black continuous curve is again the number density of the one-kink ground state of the sine-Gordon model. As we see from the figure this prediction works very well as long as one compares to results of sufficiently fine lattice spacing. The data for $a\sqrt{\lambda/(2\pi)}=0.03075$ can be hardly distinguished from that of $a\sqrt{\lambda/(2\pi)}=0.067$.
We checked the lattice corrections for larger values of $m$, and found them to be smaller.

As is clear from Figs.~\ref{nB1}--\ref{nB_compare_a}, the single baryon ground state completely breaks translation invariance. That this symmetry can break is a result of taking the large-$N$ limit : the energy cost of moving the baryon is proportional to $N$, and the latter is strictly infinite in our work. For finite values of $N$ the true ground state will presumably have total zero momentum and will not break translations. A finite density of baryons {\em can}, however, induce a spontaneous breakdown of translations, even at finite $N$. This will happen if the ground state contains a crystal of baryons which leaves invariant only a subgroup of the full translation symmetry. The cost of moving such a crystal scales like $(N\times {\rm Volume})$ and so will be infinite in the thermodynamical limit, for finite values of $N$ as well.\footnote{We thank Yigal Shamir for emphasizing this issue to us.}\,\footnote{In our one-dimensional case, any spontaneous breaking of continuous symmetry is removed by IR fluctuations, but these are suppressed as long as we are at large-$N$.} In any event, we work at $N=\infty$, and so in our case even a single baryon breaks translation invariance.

\begin{figure}[p]
\centerline{
\includegraphics[width=12cm]{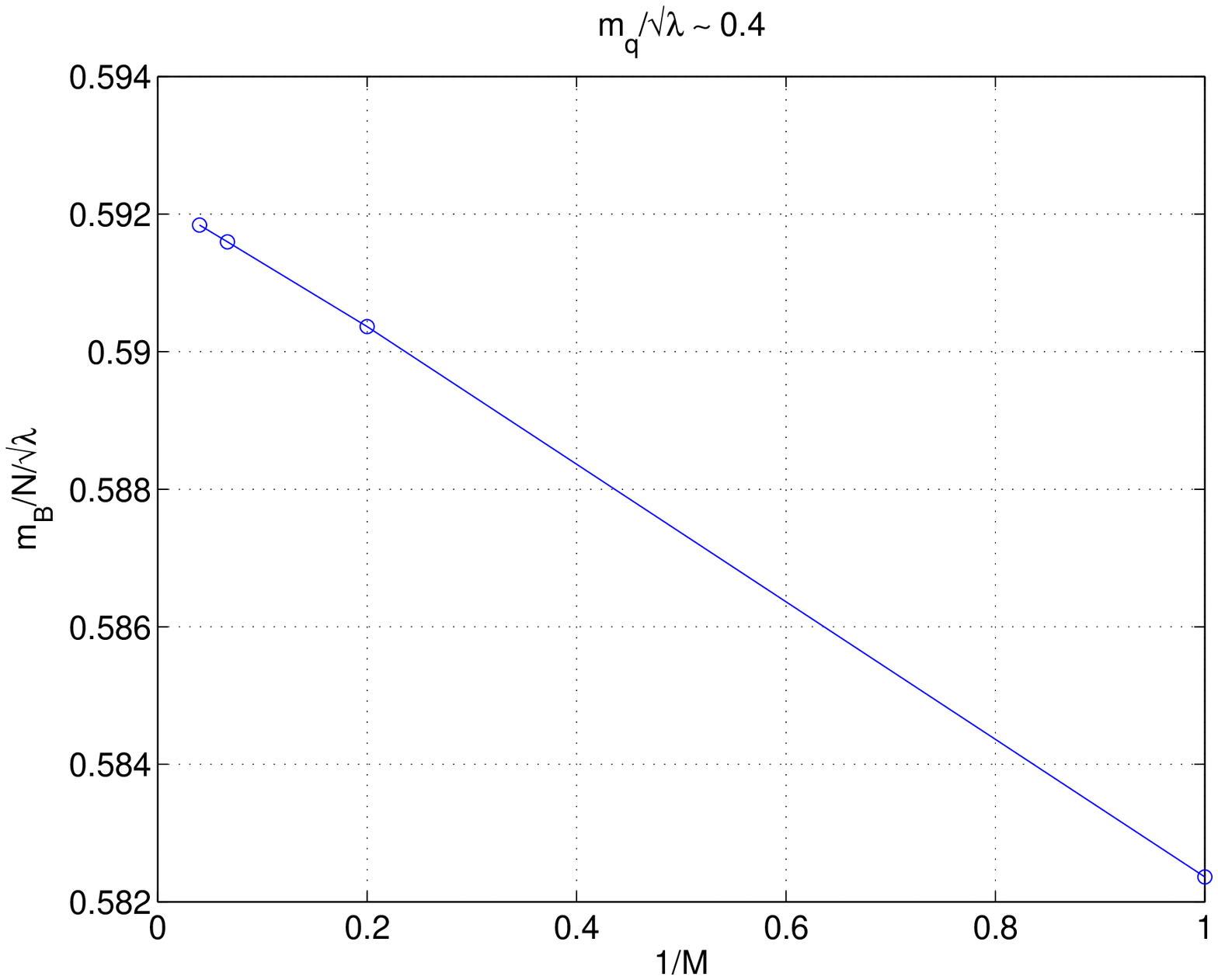}
}
\centerline{
\includegraphics[width=12cm]{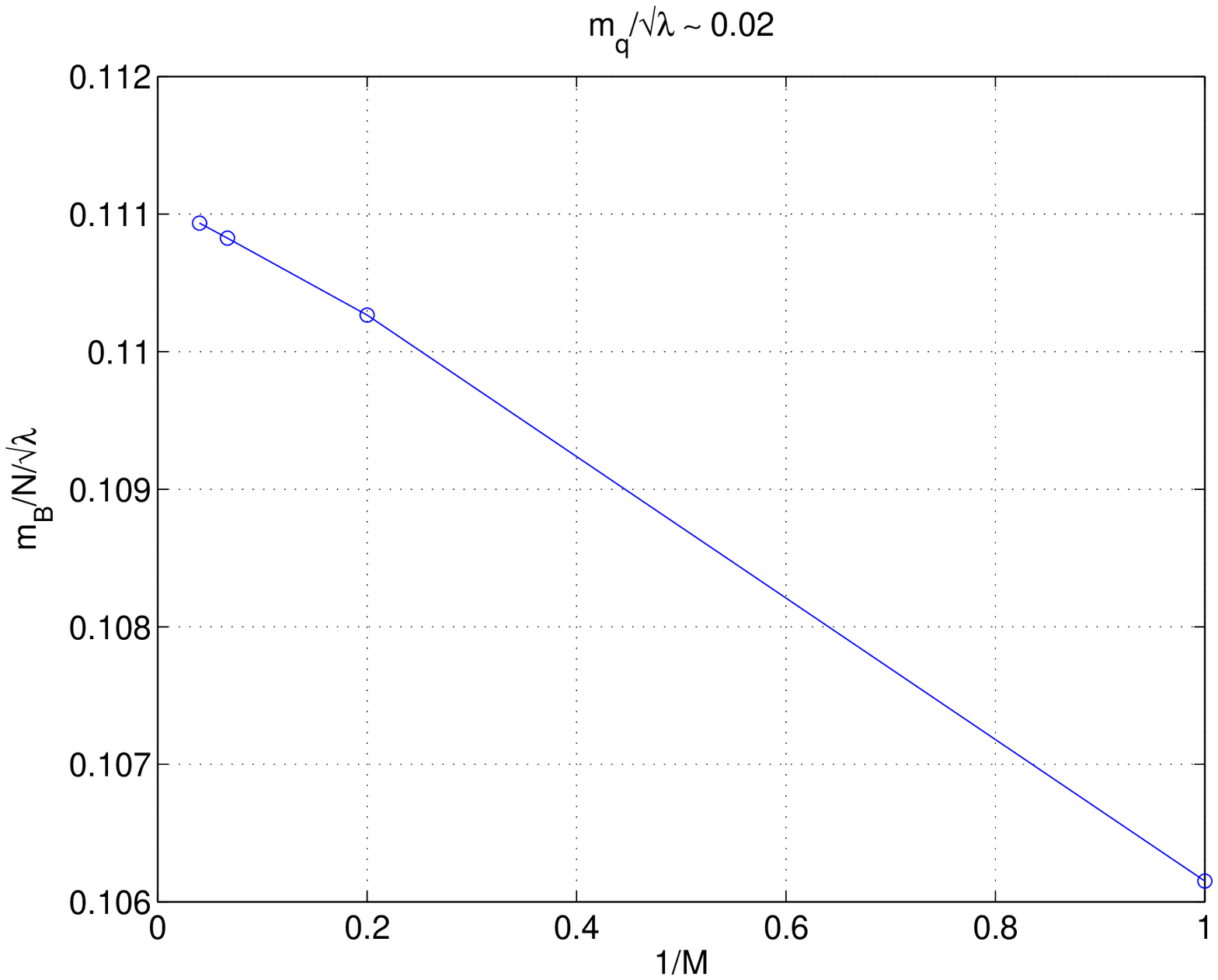}
}
\caption{Baryon mass at $a\sqrt{\lambda/(2\pi)} = 0.123$ and $L\sqrt{\lambda/(2\pi)}=16$ as a function of $1/M$. \underline{Top panel:} An intermediate quark mass; $m/\sqrt{\lambda}=1/\sqrt{2\pi}$. \underline{Bottom panel:} Small quark mass; $m/\sqrt{\lambda}=0.05/\sqrt{2\pi}$. }
\label{m_B}
\end{figure}

We proceed by analyzing the baryon mass $m_B$ which we calculate by subtracting the energy of the $B=0$ ground state from the energy of the $B=1$ ground state. Again, this was first done in Ref.~\cite{Salcedo}, but by using $M=1$. Here we wish to check whether the volume chosen there was large enough for the $1/M$ corrections to be negligible. The results are presented in Fig.~\ref{m_B} where we see that for $L\sqrt{\lambda/(2\pi)}=16$ and $a\sqrt{\lambda/(2\pi)}=0.123$, the $1/M$ corrections to $m_B/(N\sqrt{\lambda})$ are of $\sim O(5\,\%)$ for a light quark mass of $m/\sqrt{\lambda}\simeq 0.02$, and around $\sim O(1\,\%-2\,\%)$ for heavy quark masses of $m/\sqrt{\lambda}\simeq 0.4$. As a comparison, for the parameters chosen in the lower panel of Fig.~\ref{m_B}, Ref.~\cite{Salcedo} obtained $m_B/(N\sqrt{\lambda}) \simeq 0.579$, while we see that the large-$M$ extrapolation results in $m_B/(N\sqrt{\lambda}) \simeq 0.592$.

\subsection{Further comparisons with the sine-Gordon soliton}
\label{comp_SG}

As mentioned above, for light quarks obeying $m/\sqrt{\lambda} \ll 1$ the continuum version of the classical Hamiltonian ${\cal H}_F$ reduces, for a certain ansatz for $\rho^a_{xy}$, to the Hamiltonian of the sine-Gordon model \cite{Salcedo}. For $B=1$ this ansatz is the sine-Gordon soliton and in this subsection we wish to check our calculations further by comparing the energy and height of the sine-Gordon soliton with $m_B$ and with the maximum of the baryon density $\rho_{\rm max}$. 

\begin{figure}[hbt]
\centerline{
\includegraphics[width=12.5cm]{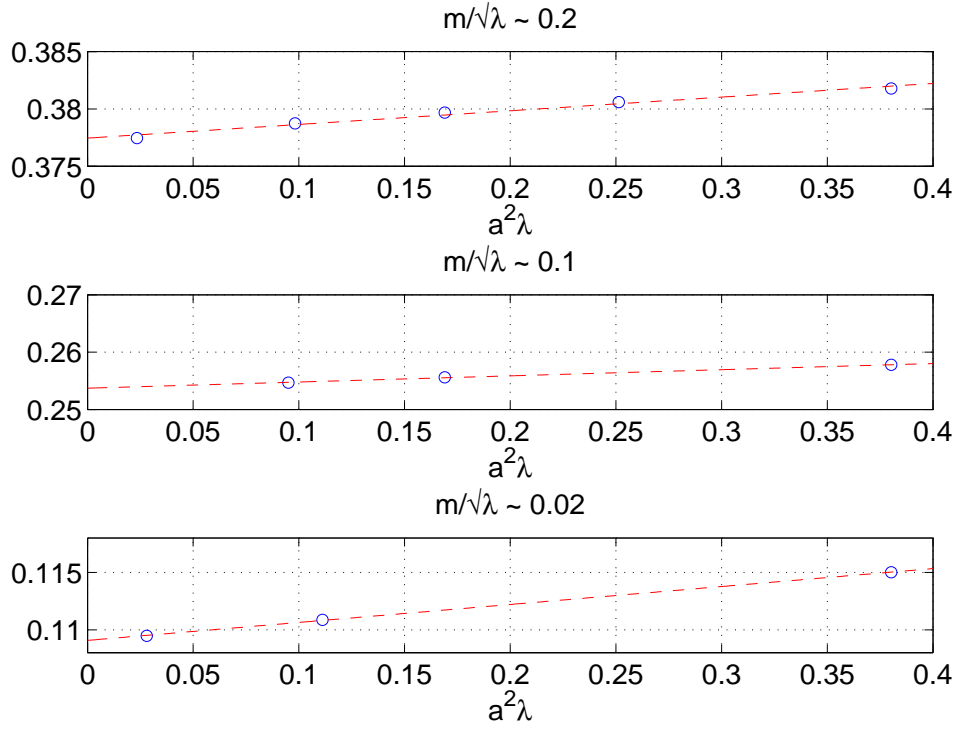}
}
\caption{Continuum extrapolations for $m_B/(N\sqrt{\lambda})$ and different quark masses for $M=15$.}
\label{mB_cont}
\end{figure}

\begin{figure}[hbt]
\centerline{
\includegraphics[width=12.5cm]{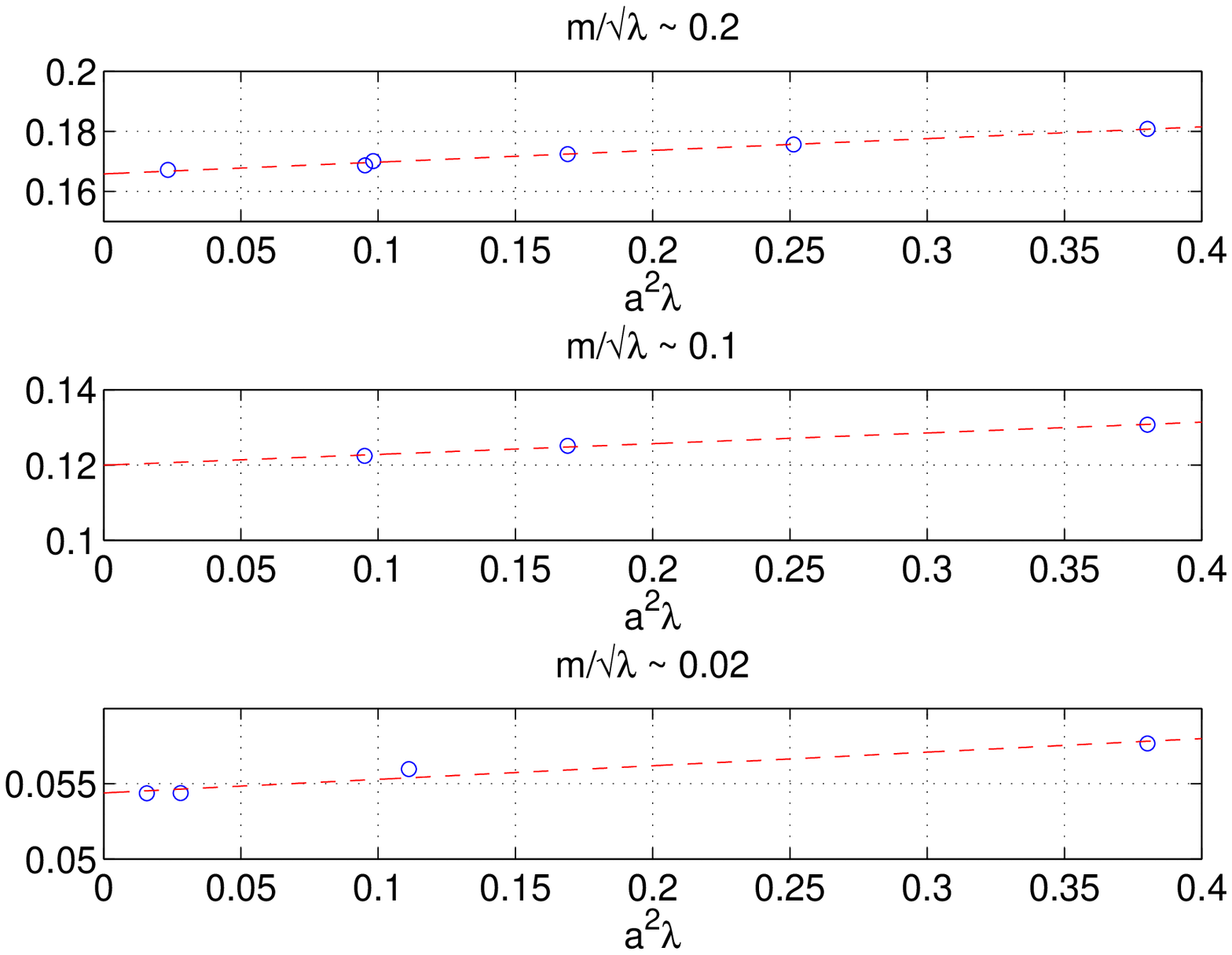}
}
\caption{Same as Fig.~\ref{mB_cont} but for $\left({\rm Max} \<\psi^\dag \psi\>\right)/(N\surd\lambda)$.}
\label{rmax_cont}
\end{figure}

\begin{figure}[p]
\centerline{
\includegraphics[width=10cm]{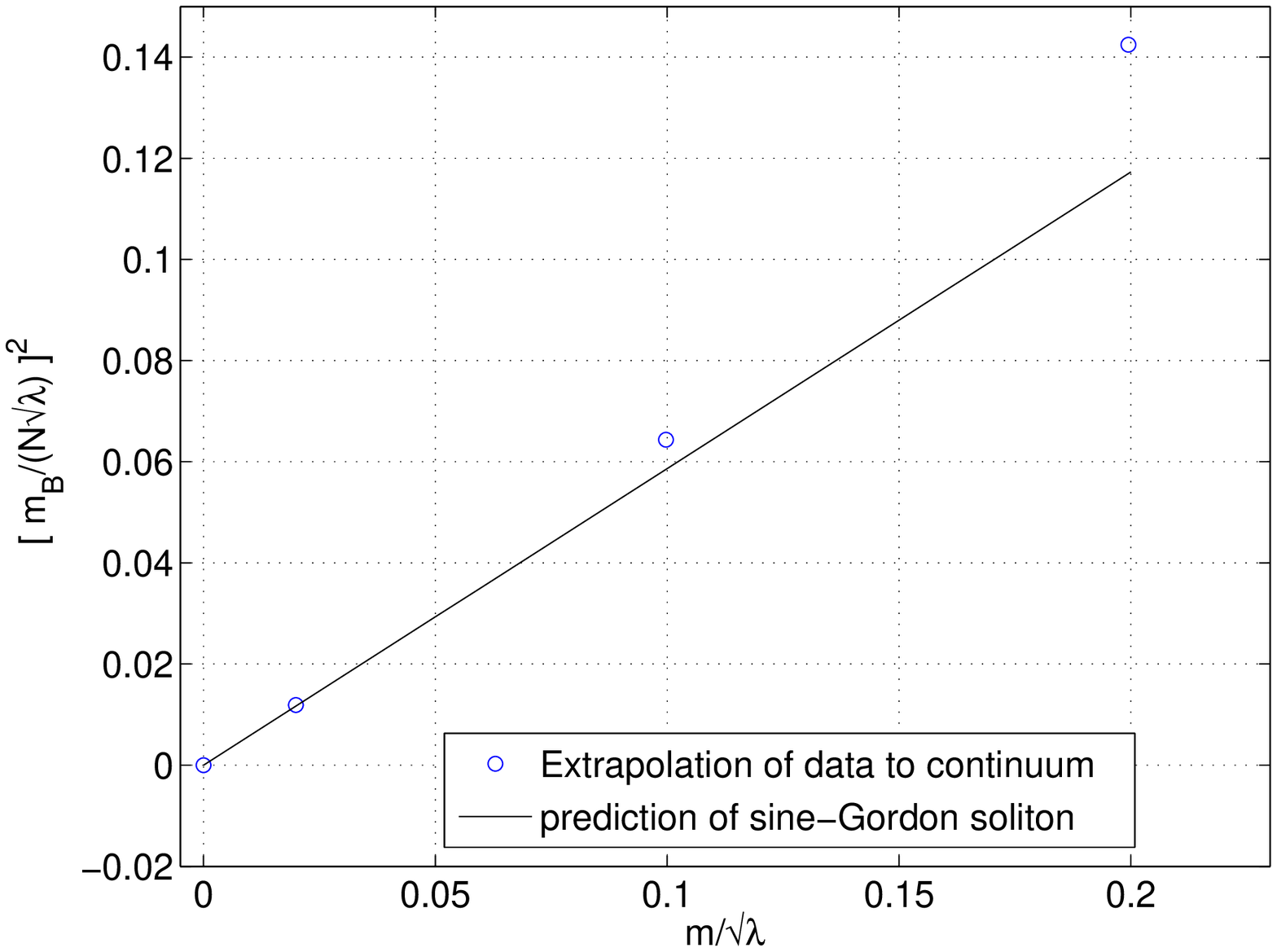}\quad
\includegraphics[width=10cm]{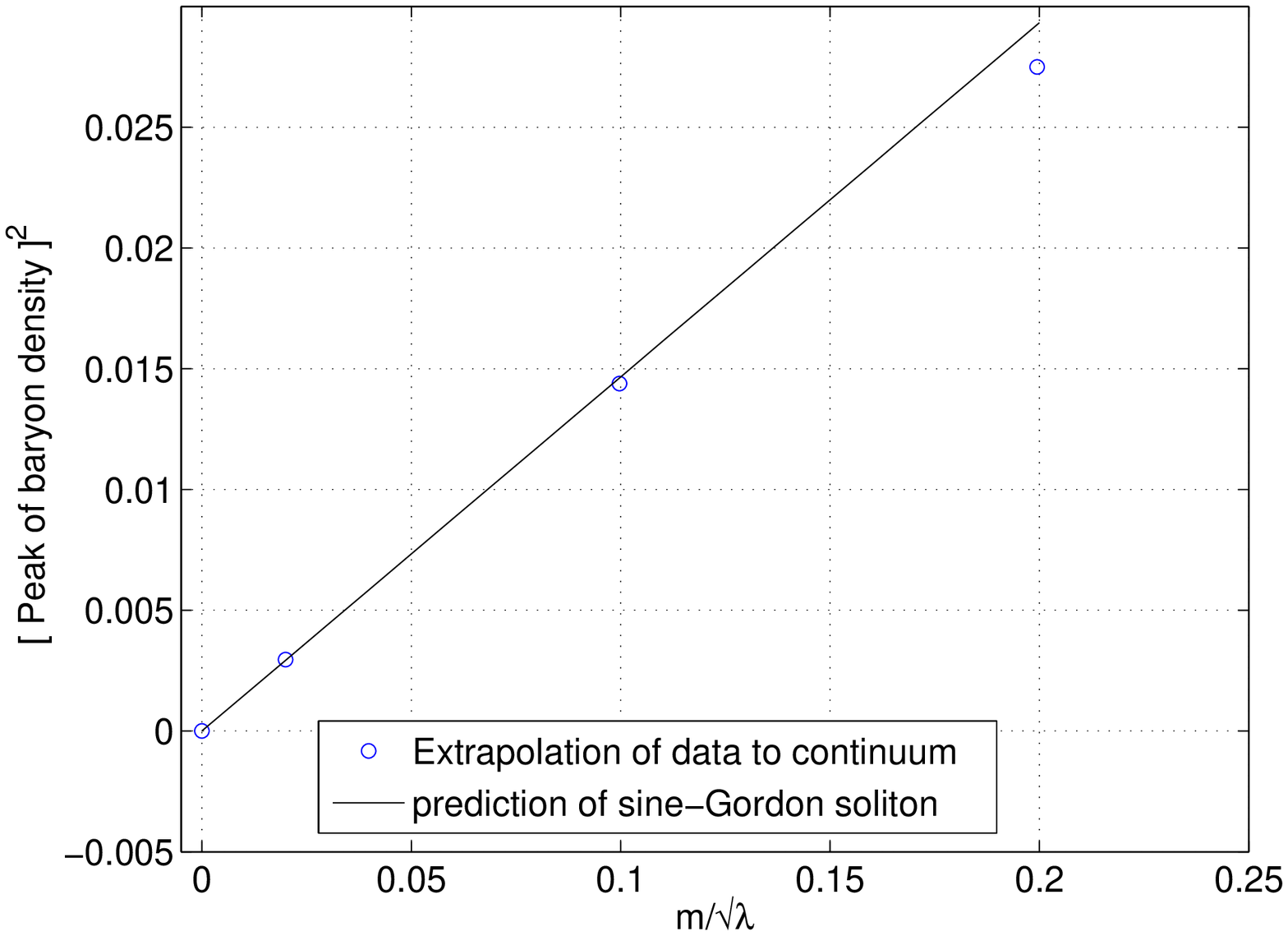}
}
\caption{Comparison of the results obtained from the continuum extrapolation appearing in Fig.~\ref{mB_cont} and Fig.~\ref{rmax_cont} to the predictions of the sine-Gordon soliton.}
\label{mBrmax_vs_SG}
\end{figure}

We begin by performing continuum extrapolations for $m_B$ and $\rho_{\rm max}$ obtained for three values of the quark mass : $m/\sqrt{\lambda}=(0.5,0.25,0.05)/\sqrt{2\pi}$. These results were calculated using $M=15$ and $L\sqrt{\lambda/(2\pi)}=16, 24, 48$ respectively, and are presented in Figs.~\ref{mB_cont} and~\ref{rmax_cont}.
In Fig.~\ref{mBrmax_vs_SG} we compare the continuum values of $m_B$ and $\rho_{\rm max}$ to the following predictions of the sine-Gordon model.
\begin{equation}
\left(\frac{m_B}{N\surd\lambda}\right)^2 = \frac{8}{\sqrt{6\pi^{3}}}\, \frac{m}{\surd\lambda}, \qquad \left(\frac{\<\psi^\dag \psi\>_{\rm max}}{N\surd\lambda}\right)^2 = \frac{2}{\sqrt{6\pi^3}} \frac{m}{\surd\lambda}.
\end{equation}
As the figure shows this prediction agrees with our data for small enough masses. Specifically, this means that the baryon mass and the inverse of its width, decrease with the quark mass to zero, and that that they so approximately like the Sine-Gordon model predicts, i.e. linear in $\surd m$. Thus, at small masses, the baryon turns into a delocalized massless excitation of the vacuum. This is very different than the way the four-dimensional baryon behaves in the chiral limit, and is precisely the reason we wanted to solve the `t Hooft model for general quark masses.

\section{Finite density and the equation of state}
\label{n_B}

\begin{figure}[p]
\centerline{
\includegraphics[width=10cm]{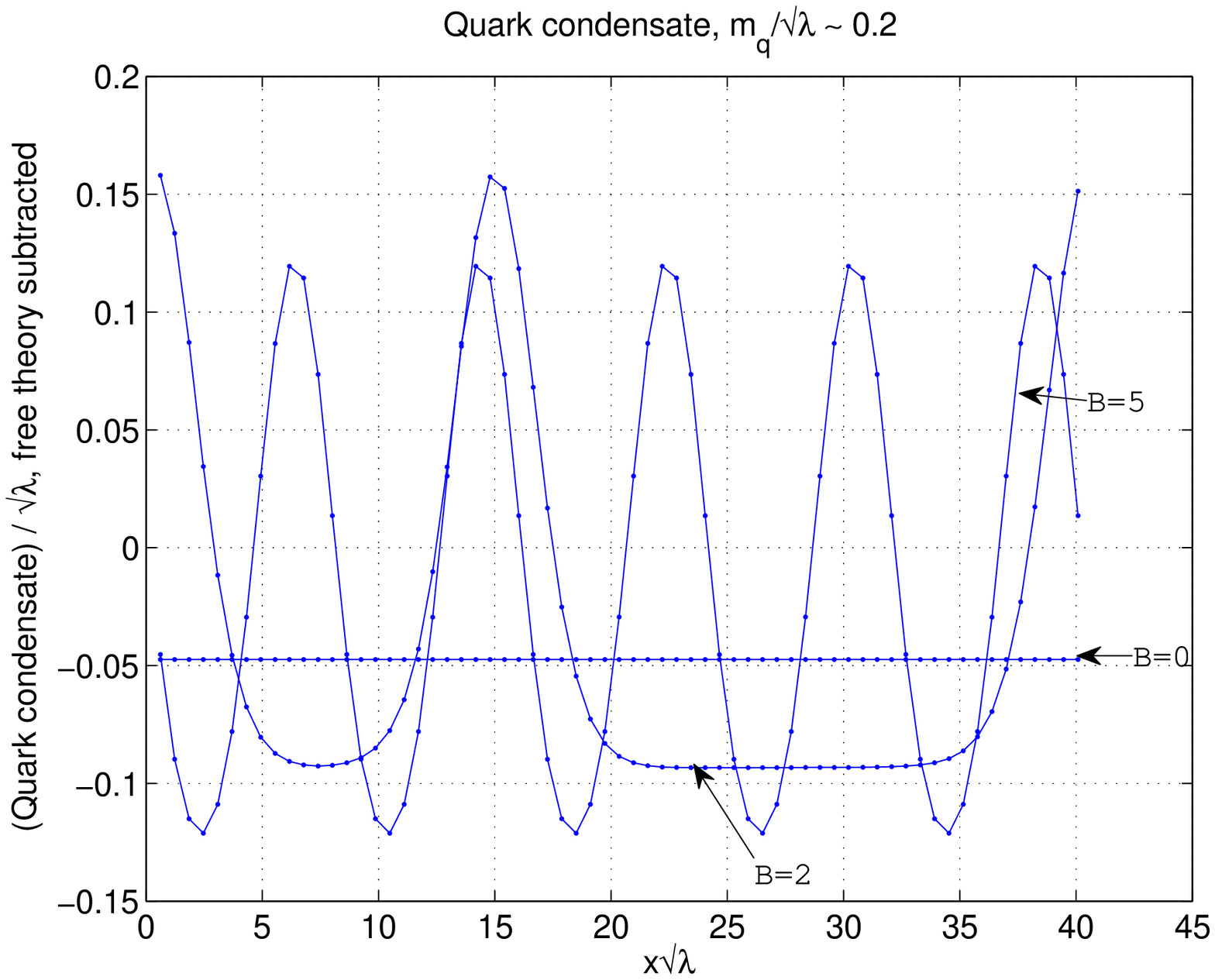}\qquad
\includegraphics[width=10cm]{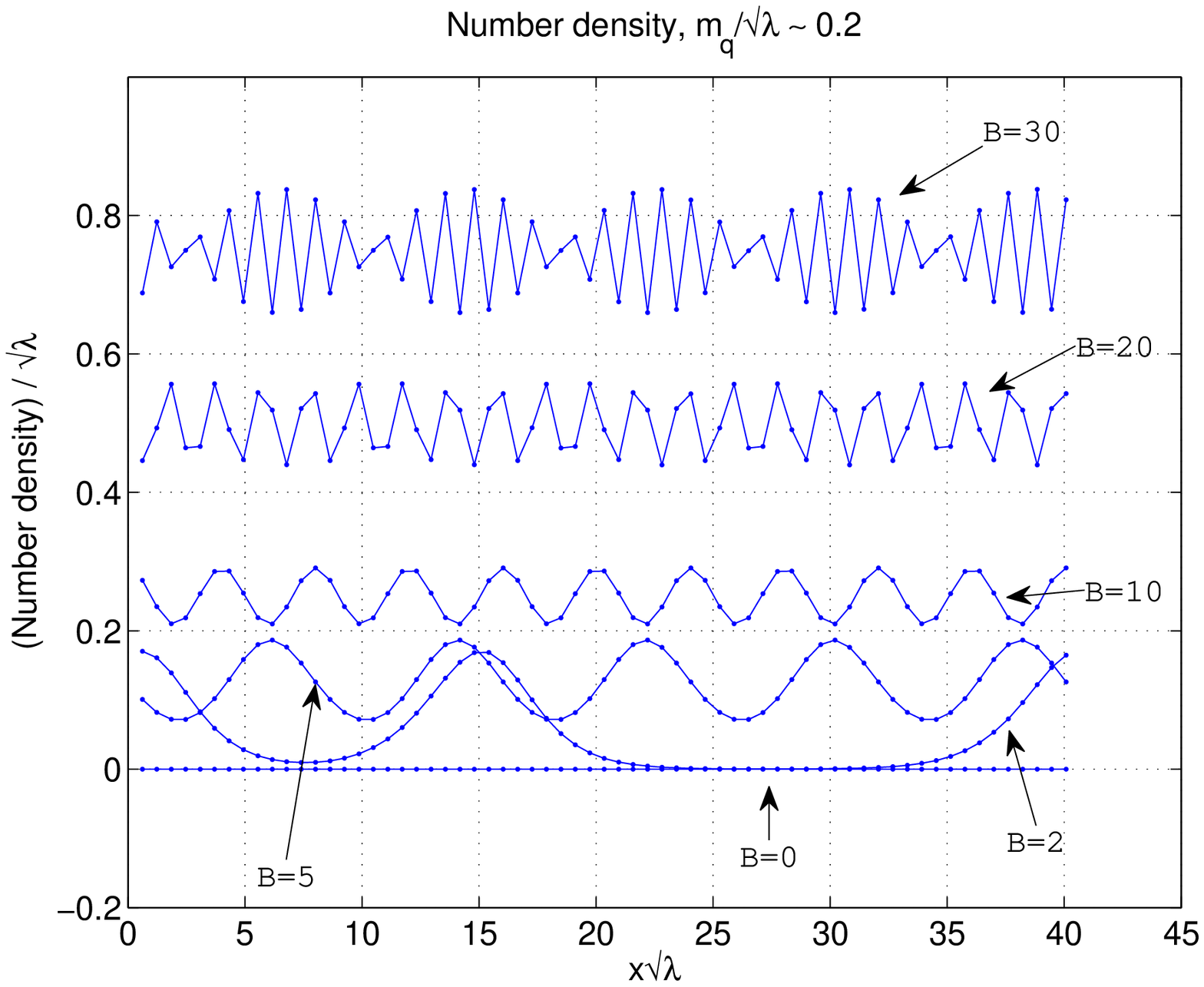}
}
\caption{Expectation values of $\<\bar \psi \psi\>/N$ (left panel) and the density $\<\psi^\dag \psi\>/N$ (right panel) as a function of the spatial coordinate $x$. Here $a\sqrt{\lambda/(2\pi)}=0.123$, $m/\sqrt{\lambda}=0.5/\sqrt{2\pi}$, $L\sqrt{\lambda/(2\pi)}=16$, and $M=25$. The baryon number showed are $B=0,3,5,10,15,20,25,30$.
}
\label{pbp_nB_1}
\end{figure}

\begin{figure}[p]
\centerline{
\includegraphics[width=10cm]{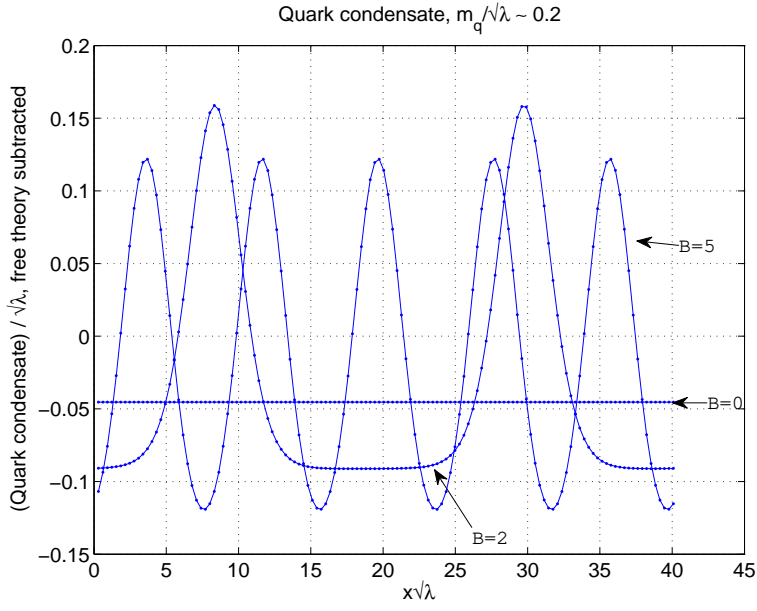}\qquad
\includegraphics[width=10cm]{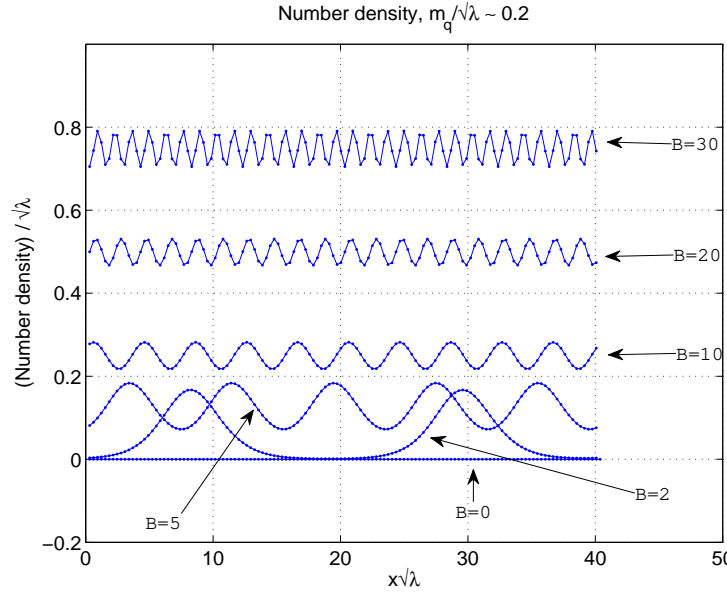}
}
\caption{Same as Fig.~(\ref{pbp_nB_1}), but for a finer lattice of $a\sqrt{\lambda/(2\pi)}=0.0615$ and $M=15$.
}
\label{pbp_nB_2}
\end{figure}

\begin{figure}[p]
\centerline{
\includegraphics[width=10cm]{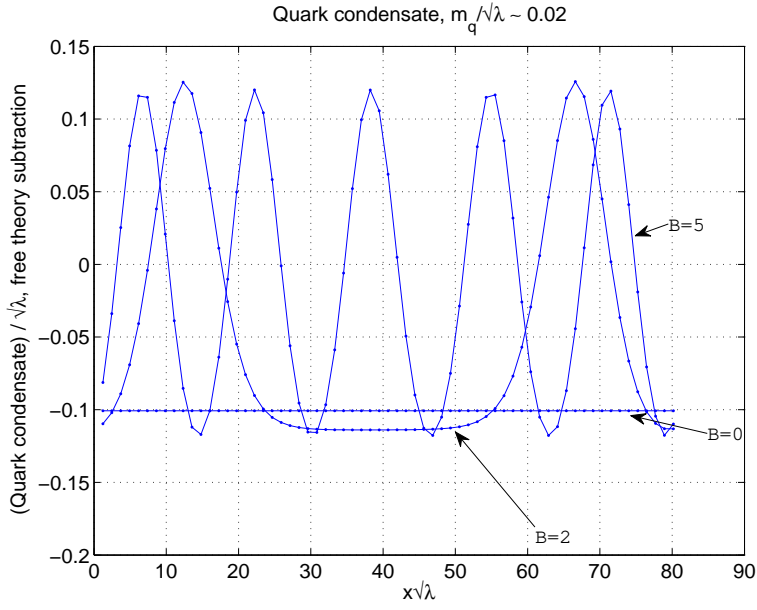}\qquad
\includegraphics[width=10cm]{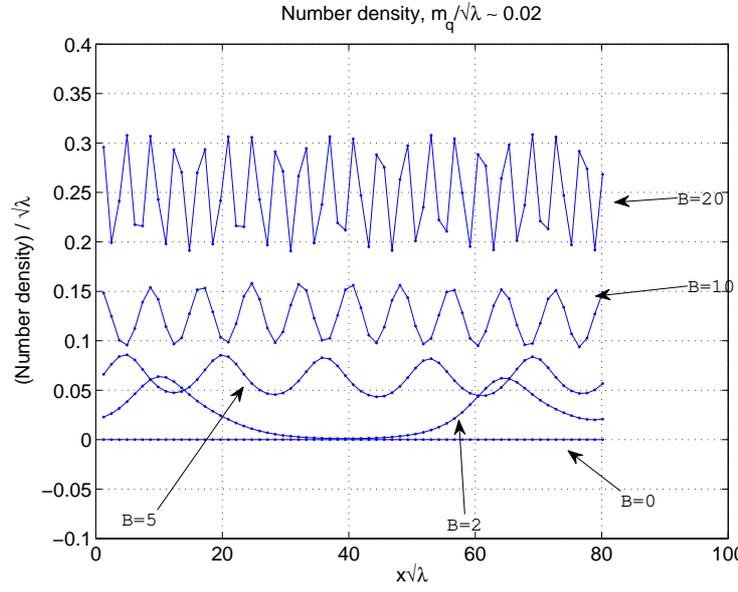}
}
\caption{Same as Fig.~(\ref{pbp_nB_1}), but for $a\sqrt{\lambda/(2\pi)}=0.246$, $m/\sqrt{\lambda}=0.05/\sqrt{2\pi}$, $L\sqrt{\lambda/(2\pi)}=32$, and $M=15$. Note the different scale in the x-axis compared to the one of Figs.~\ref{pbp_nB_1}--\ref{pbp_nB_2}}
\label{pbp_nB_3}
\end{figure}

This section contains the main results of our study: the ground state of a system at nonzero baryon density. The densities we study are obtained by putting a large baryon number of $0\le B\le 30$ in a box of fixed size $L$. We begin by presenting, in Figs.~\ref{pbp_nB_1}--\ref{pbp_nB_3}, the results of the quark condensate $\<\bar \psi \psi\>/N$ and the density $\<\psi^\dag \psi\>/N$ as a function of the spatial coordinate for different values of the quark mass, the lattice spacing, the volume, and $B$. What we find is that for any nonzero value of $B$, translation symmetry is spontaneously broken and that at large enough values of $B/L$, a crystal is formed. 

The crystal that we find is a direct result of the repulsion between the baryons. This repulsion was mentioned as an aside in Ref.~\cite{Salcedo}, and in this paper we show it explicitly when we discuss the equation of state (see the next subsection). In higher dimensions one also expects that a formation of a crystal, as long as $N$ is infinite \cite{Son}. To understand this think about the baryons as classical interacting particles (rather than emergent coherent objects that do not fluctuate). In this picture a baryon-baryon potential can  either be repulsive or attractive, depending on the dynamics. In the former case (which is what happens in the single flavor $1+1$ case, as we show in our paper), putting a bunch of baryons in a box obviously creates a crystal. What happens in four dimensions? physical, $3$-color, baryons have long-distance attraction in four dimensions, but there is still a hard-core repulsion at short distances. Thus, the distance at which the baryon-baryon potential has a minimum, is a finite, $O(\Lambda_{\rm QCD})$ value, which we denote by $r_0$. If this picture survives the large-$N$ limit, then the absence of fluctuations at that limit tells us that the inter-baryon distance will exactly be $r_0$, which means that they will form a crystal at infinite volume.

Naturally, the crystal structure is contaminated with lattice artifacts when its wave length is close to the lattice spacing (see the odd looking structure of the crystal for $B=30$ in the right panel of Fig.~\ref{pbp_nB_1}). These artifacts go away when we either decrease the density (see $B=10$ in the same figure) or decrease the lattice spacing (see $B=30$ in the right panel of Fig.~\ref{pbp_nB_2}). This phenomenon, of increasing lattice artifacts with increasing density, is of course anticipated in advance, and is usually referred to as `lattice saturation': it happens when the number of baryons per site is close to $1$, which is the maximal number that Pauli-exclusion principle allows for a theory with a single flavor.

\begin{figure}[p]
\centerline{
\includegraphics[width=20cm]{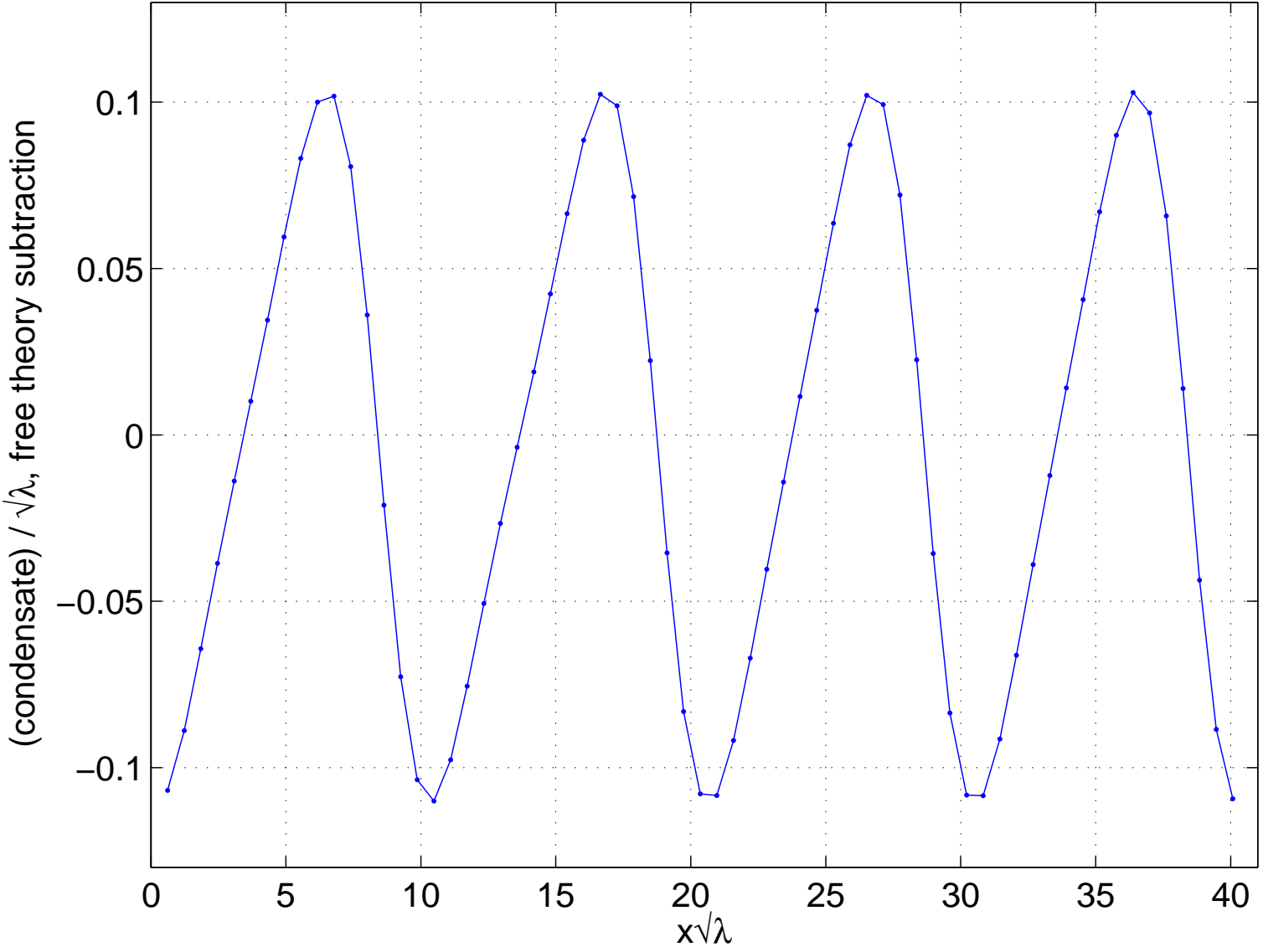}
}
\caption{The condensate $\<\bar \psi i\gamma_5 \psi\>/N$for $B=4$ and an intermediate quark mass of $m/\sqrt{\lambda}=0.5/\sqrt{2\pi}$. To produce this plot we used $a\sqrt{\lambda/(2\pi)}=0.123$, $L\sqrt{\lambda/(2\pi)}=16$, and $M=25$.}
\label{g1_ig5}
\end{figure}

\begin{figure}[p]
\centerline{
\includegraphics[width=20cm]{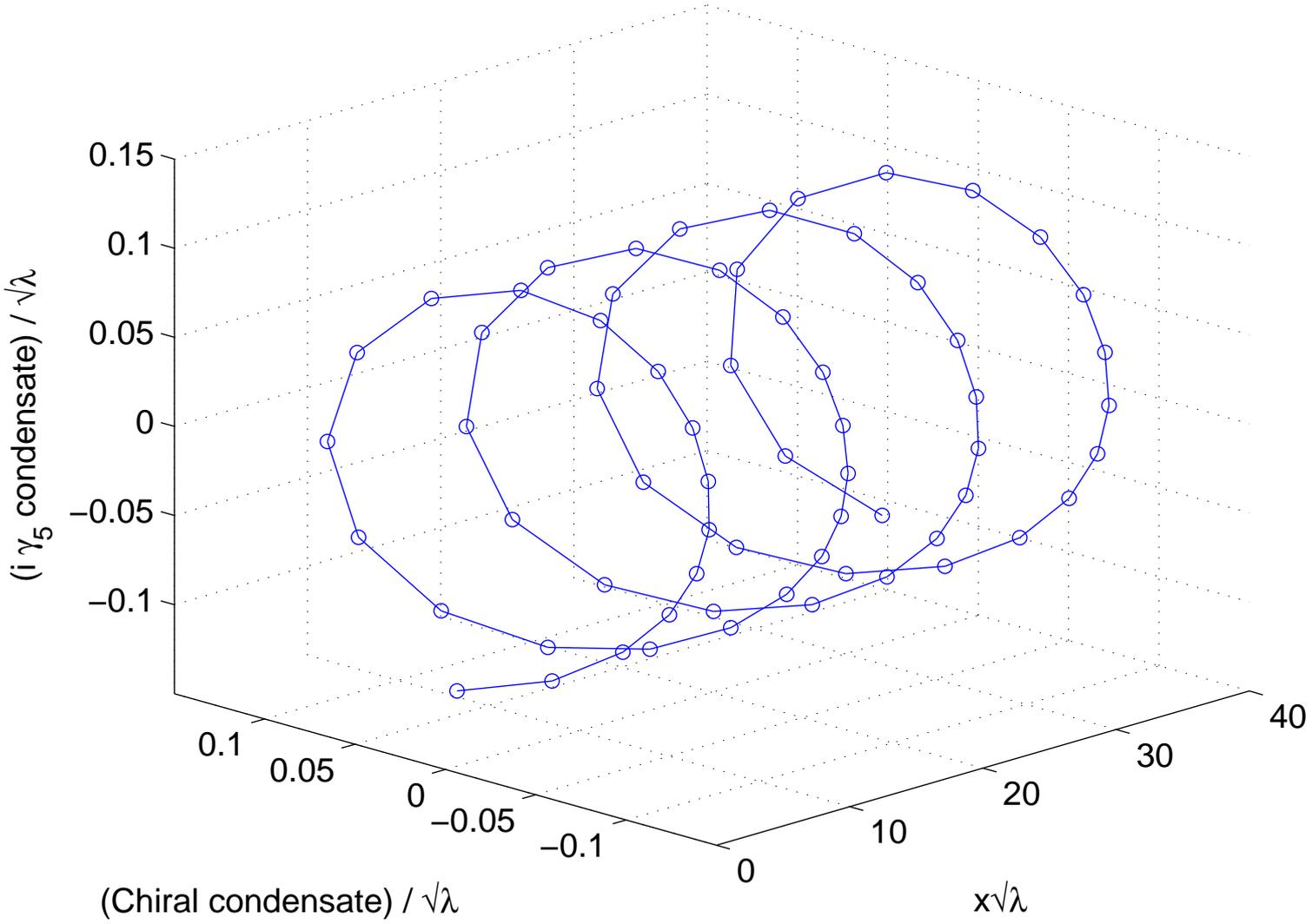}
}
\caption{The helical structure of the ground state of $B=4$ and an intermediate quark mass of $m/\sqrt{\lambda}=0.5/\sqrt{2\pi}$; $\<\bar \psi \psi\>$ on the $y$-axis, $\<\bar \psi \,i\gamma_5 \psi\>$ on the $z$ axis, and the spatial coordinate $x\sqrt{\lambda}$ on the $x$ axis. Technical parameters are the same as those of Fig.~\ref{g1_ig5}.}
\label{helic1}
\end{figure}

\begin{figure}[p]
\centerline{
\includegraphics[width=20cm]{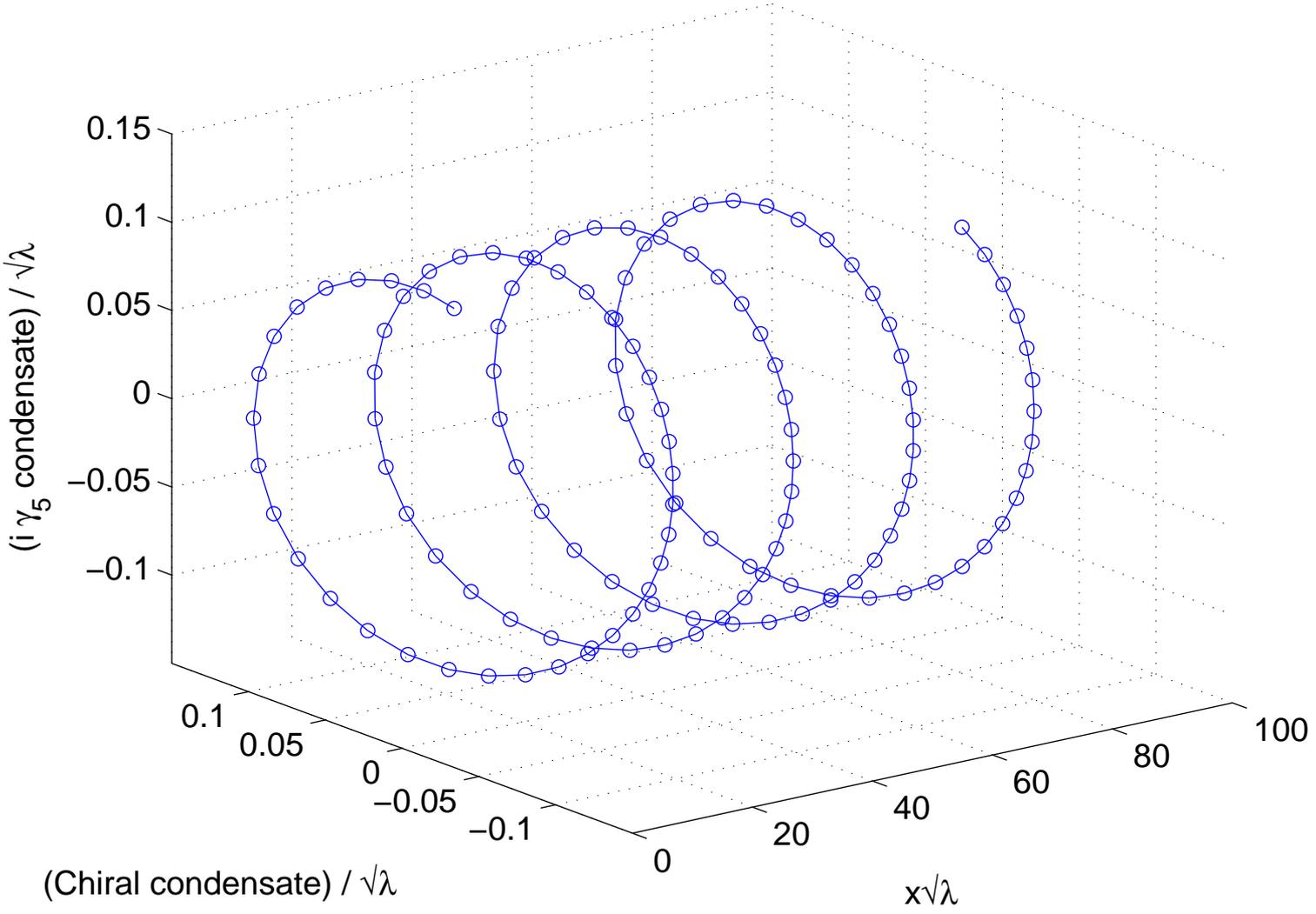}
}
\caption{Same as Fig.~(\ref{helic1}), but for $m/\sqrt{\lambda}=0.05/\sqrt{2\pi}$. Other technical parameters are $a\sqrt{\lambda/(2\pi)}=0.123$, $L\sqrt{\lambda/(2\pi)}=32$, and $M=15$. Note the different scales of the $x$, $y$ and $z$ axes compared to those of Fig.~\ref{helic1}.}
\label{helic2}
\end{figure}

\begin{figure}[p]
\centerline{
\includegraphics[width=20cm]{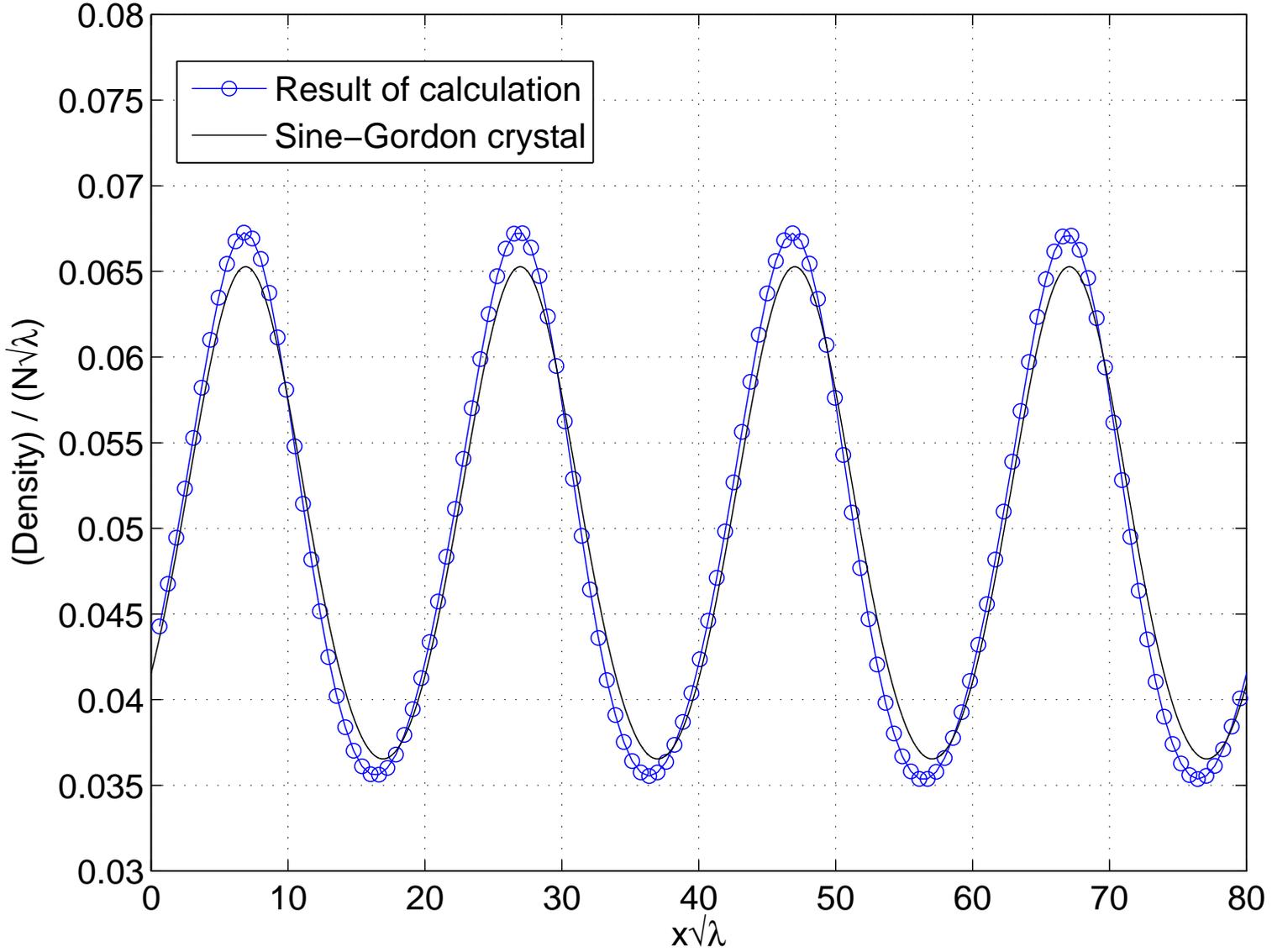}
}
\caption{The baryon density for $m/\sqrt{\lambda}=0.05/\sqrt{2\pi}$, $B=4$, $L\sqrt{\lambda/(2\pi)}=32$ and $a\sqrt{\lambda/(2\pi)}=0.123$. The solid (black) line is the prediction of the sine-Gordon crystal. The technical parameter $M$ was fixed to $15$.}
\label{nB4_light}
\end{figure}

\begin{figure}[p]
\centerline{
\begin{tabular}{c}
\includegraphics[width=10cm]{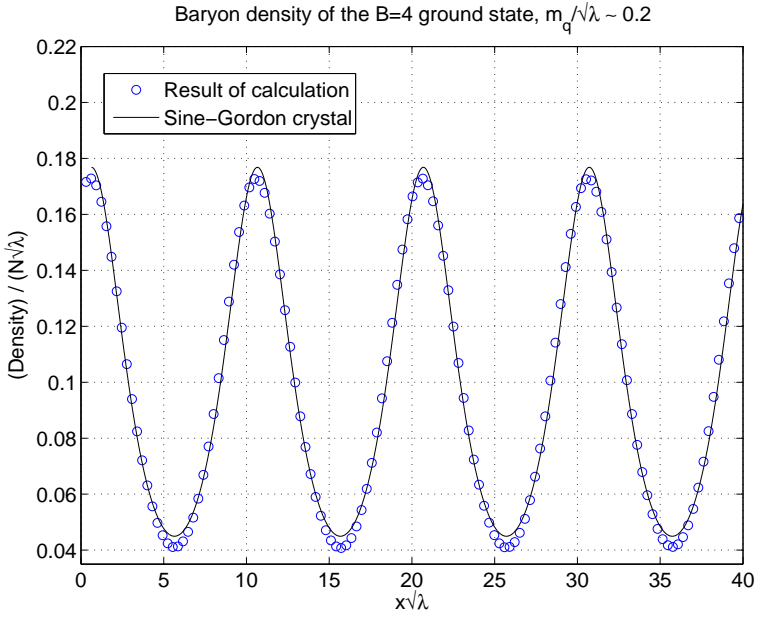}
\quad
\includegraphics[width=10cm]{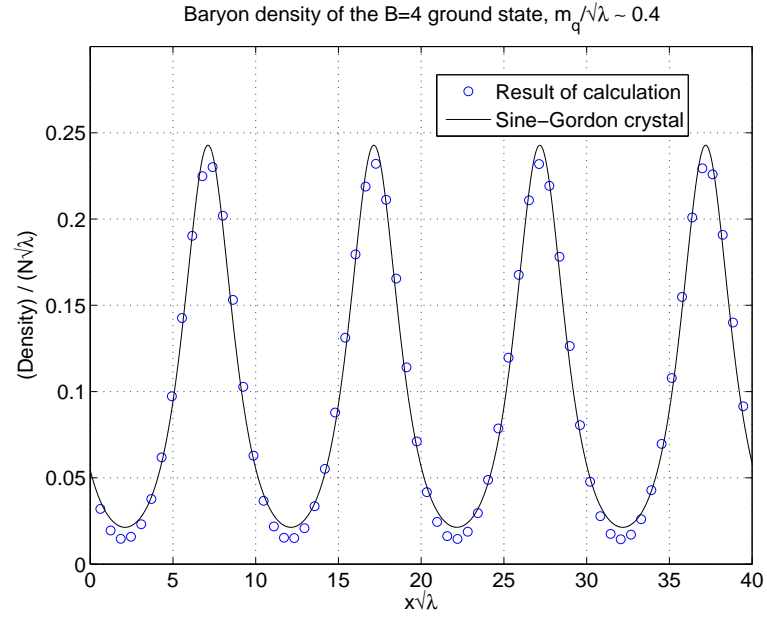}
\\ 
\includegraphics[width=10cm]{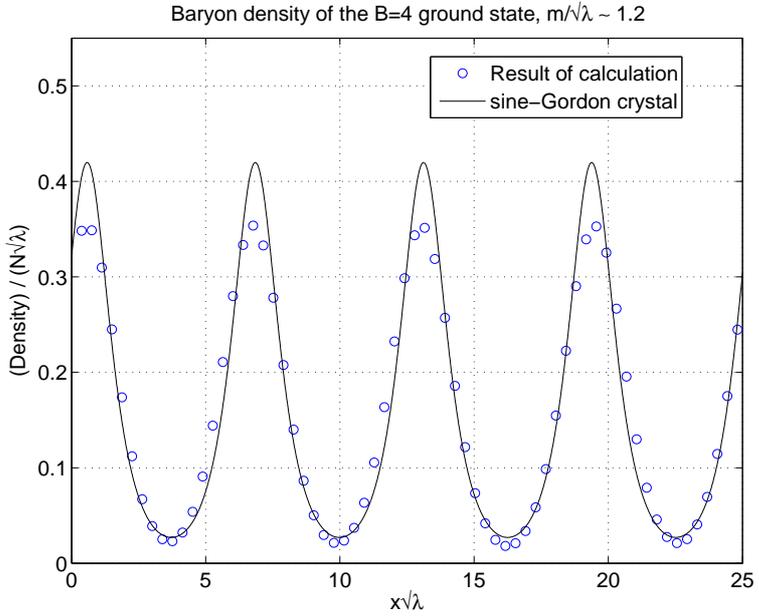}
\quad
\includegraphics[width=10cm]{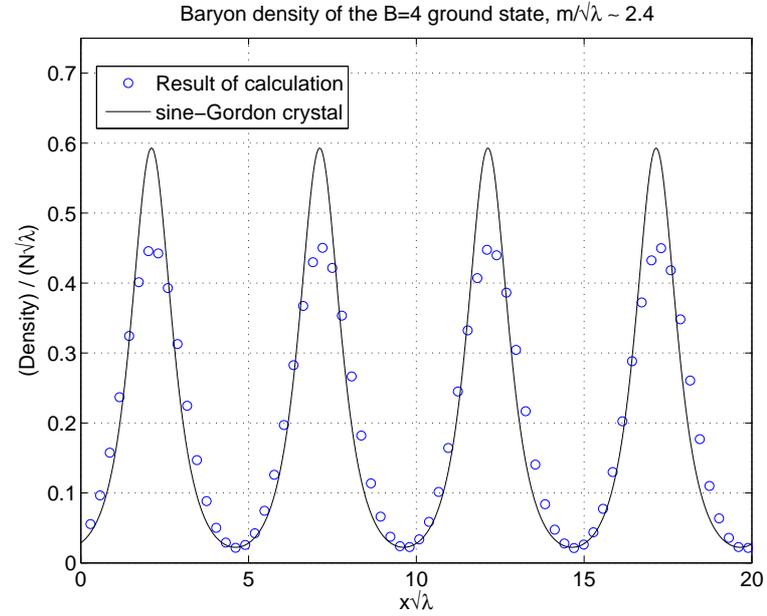}
\end{tabular}
}
\caption{The baryon density for $B=4$. \underline{Upper-left panel:} intermediately heavy quarks with $m/\sqrt{\lambda} = 0.5/\sqrt{2\pi}$ and $a\sqrt{\lambda/(2\pi)}=0.0615$. \underline{Upper-right panel:} heavy quarks with $m/\sqrt{\lambda} = 1.0/\sqrt{2\pi}$ and $a\sqrt{\lambda/(2\pi)}=0.123$. \underline{Lower-left panel:} heavier quarks with $m/\sqrt{\lambda} = 3.0/\sqrt{2\pi}$ and $a\sqrt{\lambda/(2\pi)}=0.075$.  \underline{Lower-right panel:} even heavier quarks with $m/\sqrt{\lambda} = 6.0/\sqrt{2\pi}$ and $a\sqrt{\lambda/(2\pi)}=0.075$. The solid (black) lines are the prediction of the sine-Gordon crystal. The parameter $M$ was fixed to $25$ for the upper-left panel, and $15$ for the other panels. In units of $\sqrt{\lambda/(2\pi)}$,  the volumes were $16$ for the upper plots, $10$ for the lower-left plot and $8$ for the lower-right plot. }
\label{nB4_heavy}
\end{figure}

In Fig.~\ref{g1_ig5} we plot $\<\bar \psi \,i\gamma_{5} \psi\>/N$, and find that it is also spatially modulated.
In Fig.~\ref{helic1} we show that this crystal has a helical structure: both $\<\bar\psi \psi\>/N$ and $\<\bar \psi \,i\gamma_5 \psi\>/N$ are plotted versus the spatial coordinate. The parameters of the data plotted are the same as those used to generate Fig~\ref{g1_ig5}.
In Fig.~\ref{helic2} we repeat the plot presented in Fig.~\ref{helic1}, but for a lighter quark mass of $m/\sqrt{\lambda} = 0.05/\sqrt{2\pi}$. 
In Fig.~\ref{nB4_light} we present the baryon density for the light quark case and compare to the predictions of the sine-Gordon crystal -- see \cite{SchonThies}. The agreement is impressive.
We present similar plots for heavier quark masses of $m/\sqrt{\lambda}\simeq 0.2,0.4,1.2,2.4$ in Fig.~\ref{nB4_heavy}. The deviations from the sine-Gordon crystal are increasing with increasing $m/\surd\lambda$, but at a relatively modest rate.

\begin{figure}[p]
\centerline{
\includegraphics[width=20cm]{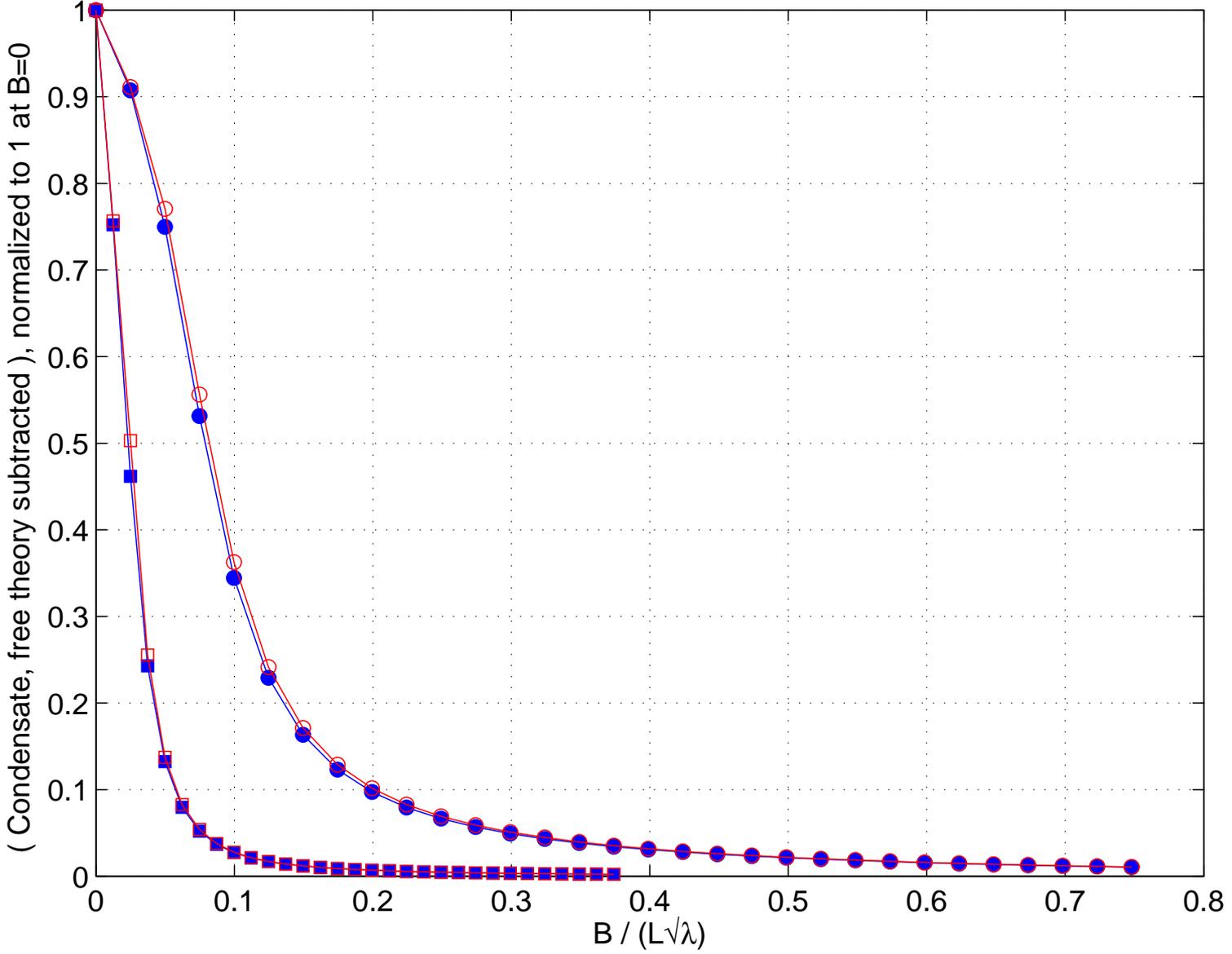}
}
\caption{The volume average of $\<\bar \psi\psi\>$ vs the baryon density $B/L$, for $a\sqrt{\lambda/(2\pi)}=0.123$ and $L\sqrt{\lambda/(2\pi)}=16$; Circles are for $m/\sqrt{\lambda}=0.5/\sqrt{2\pi}$ and squares are for $m/\sqrt{\lambda}=0.05/\sqrt{2\pi}$. For each mass we show results from both $M=1$ and $M=15$ (open red and closed blue symbols).}
\label{Avol1}
\end{figure}

\begin{figure}[p]
\centerline{
\includegraphics[width=20cm]{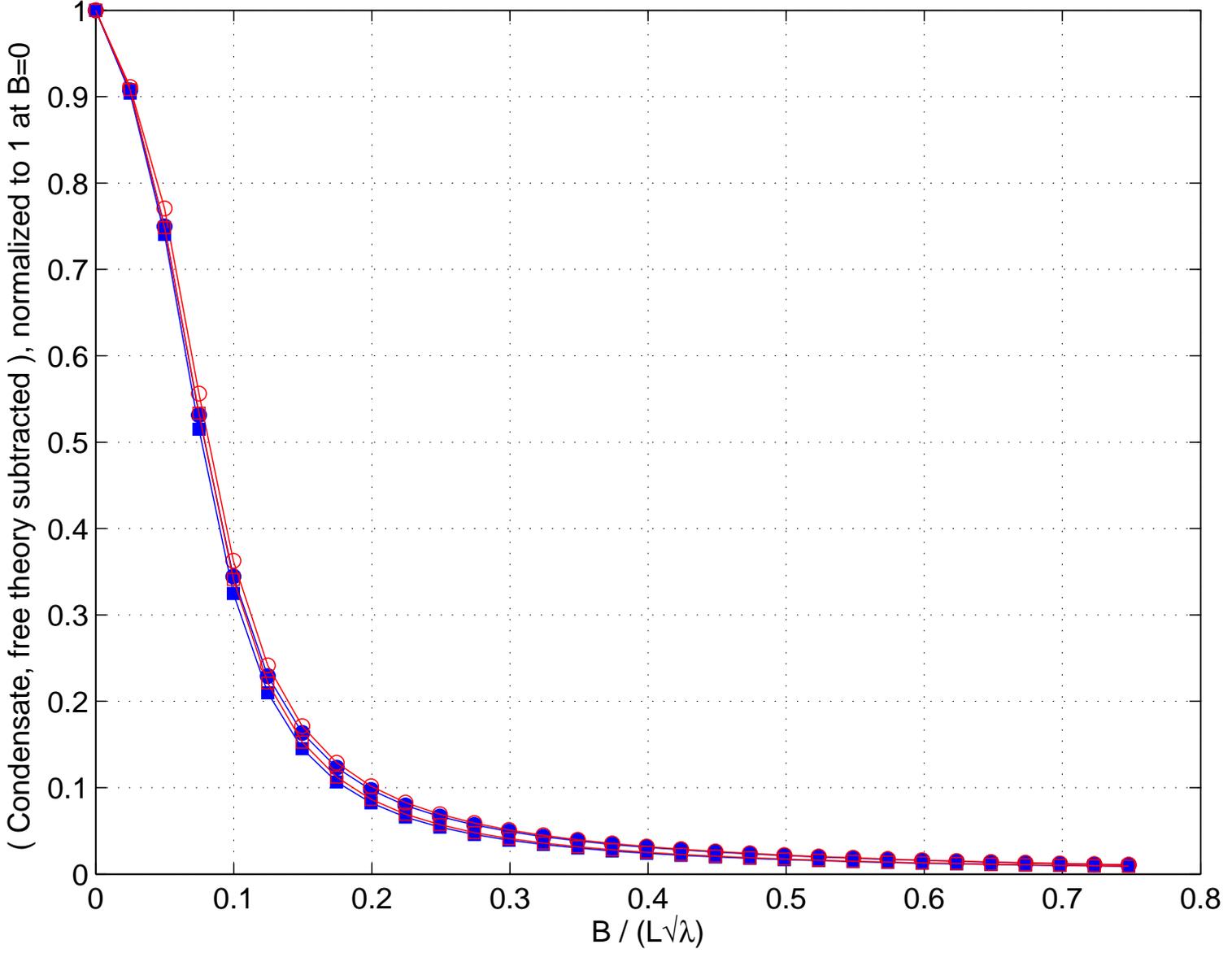}
}
\caption{The volume average of $\<\bar \psi\psi\>$ for $M=1$ (open symbols) and $M=15$ (filled symbols) and $L\sqrt{\lambda/(2\pi) }=16$; Circles are for $a\sqrt{\lambda/(2\pi)}=0.123$ and squares are for $a\sqrt{\lambda/(2\pi)}=0.0615$. Here the mass is $m/\sqrt{\lambda}=0.5/\sqrt{2\pi}$.}
\label{Avol2}
\end{figure}

Next, we integrate $\<\bar \psi \psi\>$ over the volume and present the results versus the density in Fig.~\ref{Avol1}--\ref{Avol2}. What we find is that the volume averaged value of $\<\bar \psi \psi\>$ drops with the density in a way that becomes stronger with decreasing quark mass.

\subsection{The equation of state}
\label{EoS}

\begin{figure}[p]
\centerline{
\begin{tabular}{c}
\includegraphics[height=9.5cm,width=15cm]{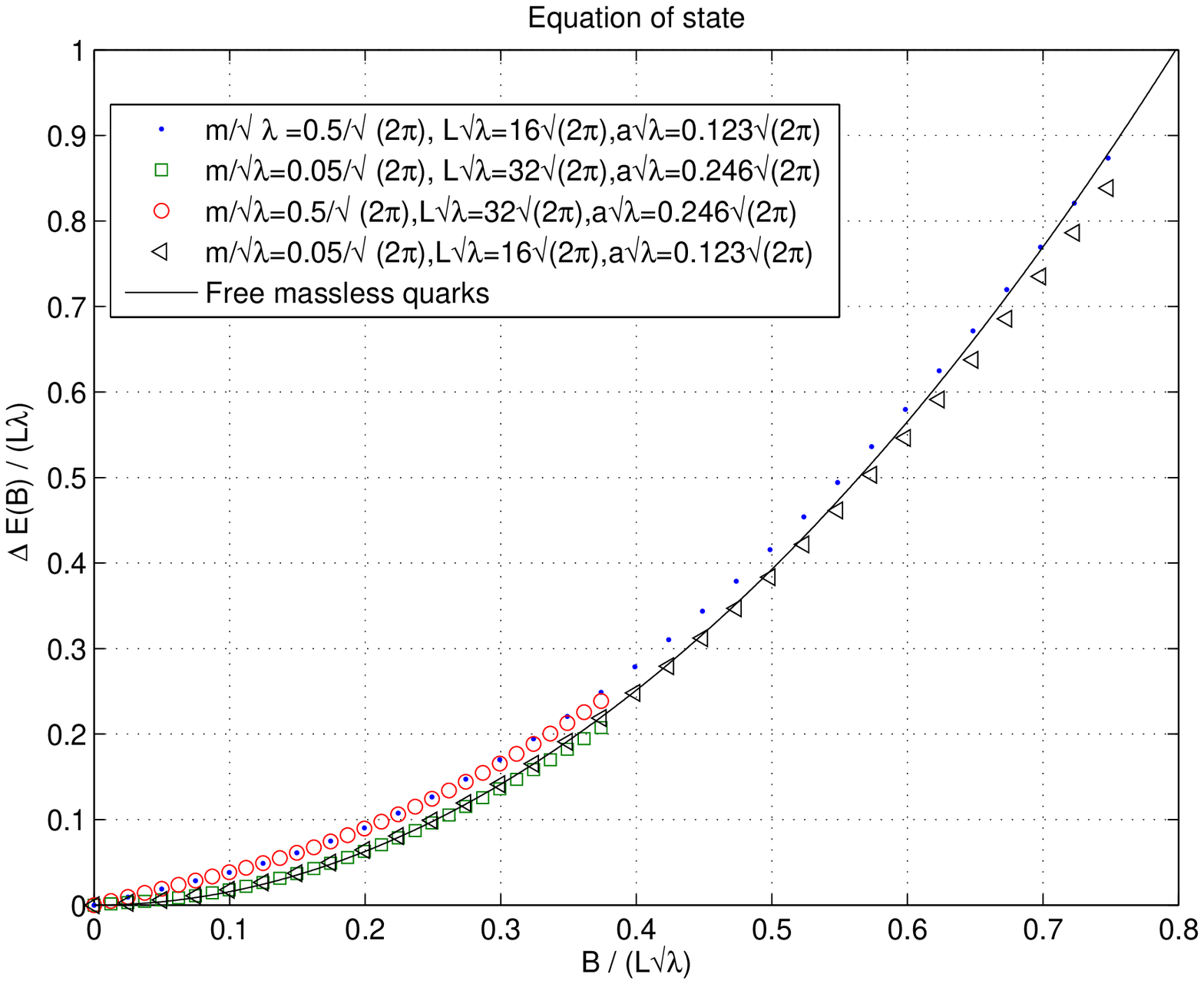}\\
\includegraphics[height=9.5cm,width=15cm]{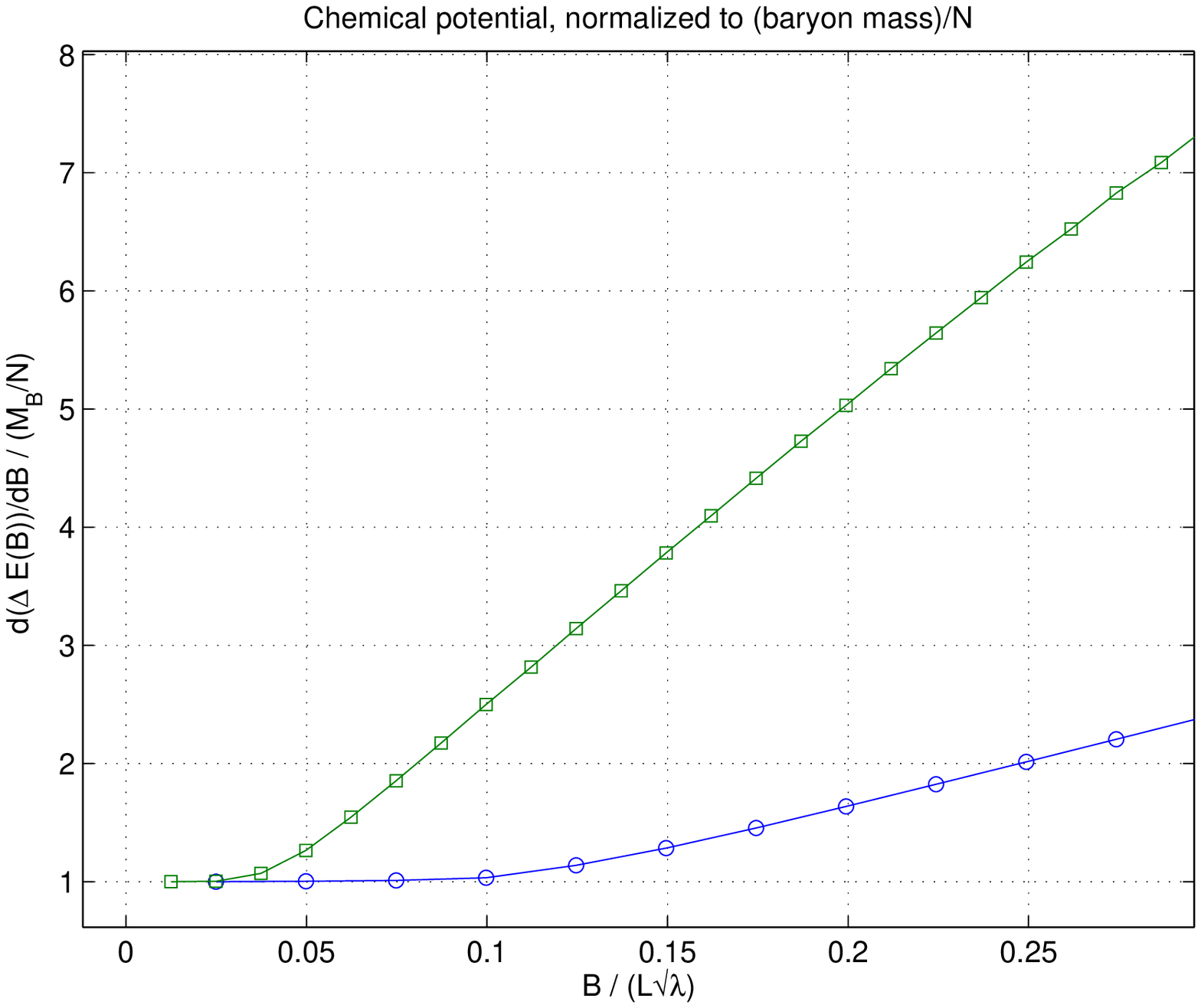}
\end{tabular}
}
\caption{\underline{Upper panel:} The $B$-dependence of the difference in energy $\Delta E(B)$ between the $B>0$ and the $B=0$ systems. \underline{Lower panel:} The chemical potential normalized to the baryon mass divided by $N$. For these calculations we choose $M=15$ and the other details of the calculations appear in the legend. The value of the baryon mass for the light ($m/\sqrt{\lambda}=0.05/\sqrt{2\pi}$)and intermediate ($m/\sqrt{\lambda}=0.5/\sqrt{2\pi}$) quark masses is $m_B/(\sqrt{\lambda} N) \simeq 0.11,0.37$, respectively.}
\label{dF}
\end{figure}

Moving on, we plot the energy $E$ of the ground state in the presence of $B$ baryons, i.e. the equation of state, in the upper panel of Fig.~\ref{dF}. Here we regularized $E$ by subtracting the $B=0$ ground state energy. We present results for both light and moderately heavy quarks (see captions). Also plotted in Fig.~\ref{dF}, as a solid (black) curve, is the equation of state for free massless quarks, in the continuum. The latter is given by 
\begin{equation}
E(B)/L = \frac\pi 2 \left(\frac{B}{L}\right)^2,
\end{equation}
and seems to be quite close to our data, especially for large enough density. This means that there is a smooth transition from Baryon physics to free quark physics as the density increases.\footnote{We thank D.~Son for pointing this issue to us.} This seems to be a nontrivial result and it would be very interesting to understand it better.

Note that the fact that the $E(B)$ is quadratic in $B$ at large-$B$ means that the would-be binding energy $\Delta$, which is given by
\begin{equation}
\Delta = \lim_{B\to \infty} \left(\frac{E(B)}{B}-E(1)\right),
\end{equation}
is proportional to $B$, is growing, and is always {\em positive}. This means that there is no binding in $1+1$ and a single flavor, or put differently that the baryons repel each other. This also means  that they would create a crystal when put together in a box, as we showed in the previous subsection.

In the lower panel of Fig.~\ref{dF} we present the numeric evaluation of $\partial E(B)/\partial B$ which is equal to the chemical potential. Here, in contrast to the upper panel, the four data sets do not fall on top of another because of the different volumes used. Note that we have normalized the $y$-axis in this plot to the constituent quark mass $m_B/N$.

\begin{figure}[p]
\centerline{
\begin{tabular}{c}
\includegraphics[width=20cm]{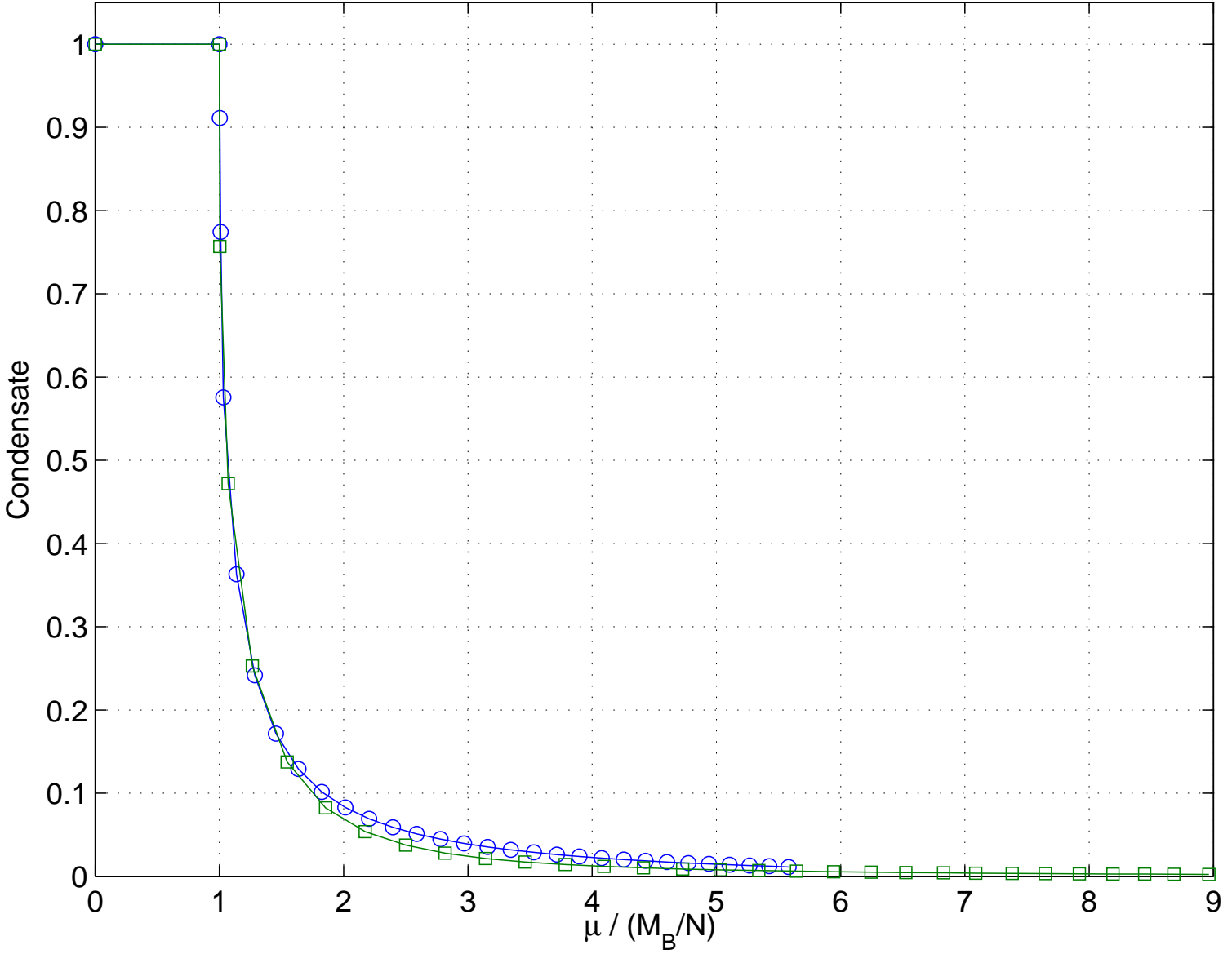}
\end{tabular}
}
\caption{The value of the quark condensate for $m/\sqrt{\lambda}=0.5/\sqrt{2\pi}, L\surd\lambda=16\surd 2\pi$, and $a\surd\lambda=0.123\surd 2\pi$ (green squares) and  $m/\sqrt{\lambda}=0.05/\sqrt{2\pi}, L\surd\lambda=32\surd 2\pi$, and $a\surd\lambda=0.246\surd 2\pi$ (blue circles). The $x$-axis is the chemical potential (normalized to  the baryon mass), and the $y$-axis is the regularized condensate, normalized to its value in the $B=0$ system.}
\label{pbp_vs_mu}
\end{figure}

Finally, in Fig.~\ref{pbp_vs_mu} we combine Figs.~\ref{Avol1},~\ref{Avol2}, and~\ref{dF}, and present the values of the regularized quark condensate, divided by its $B=0$ value, and plotted vs the chemical potential (the latter divided by $m_B/N$). The absence of data points between $\mu/(m_B/N)=0$ and $\mu/(m_B/N)=1$ reflects the absence of excitations with nonzero baryon number that are lighter than a single baryon and thus the $\mu$-independence of the ground state for $0\le \mu \le m_B/N$.

\section{Partial volume independence at large-$N$}
\label{partial}

\begin{figure}[p]
\centerline{
\includegraphics[width=11cm]{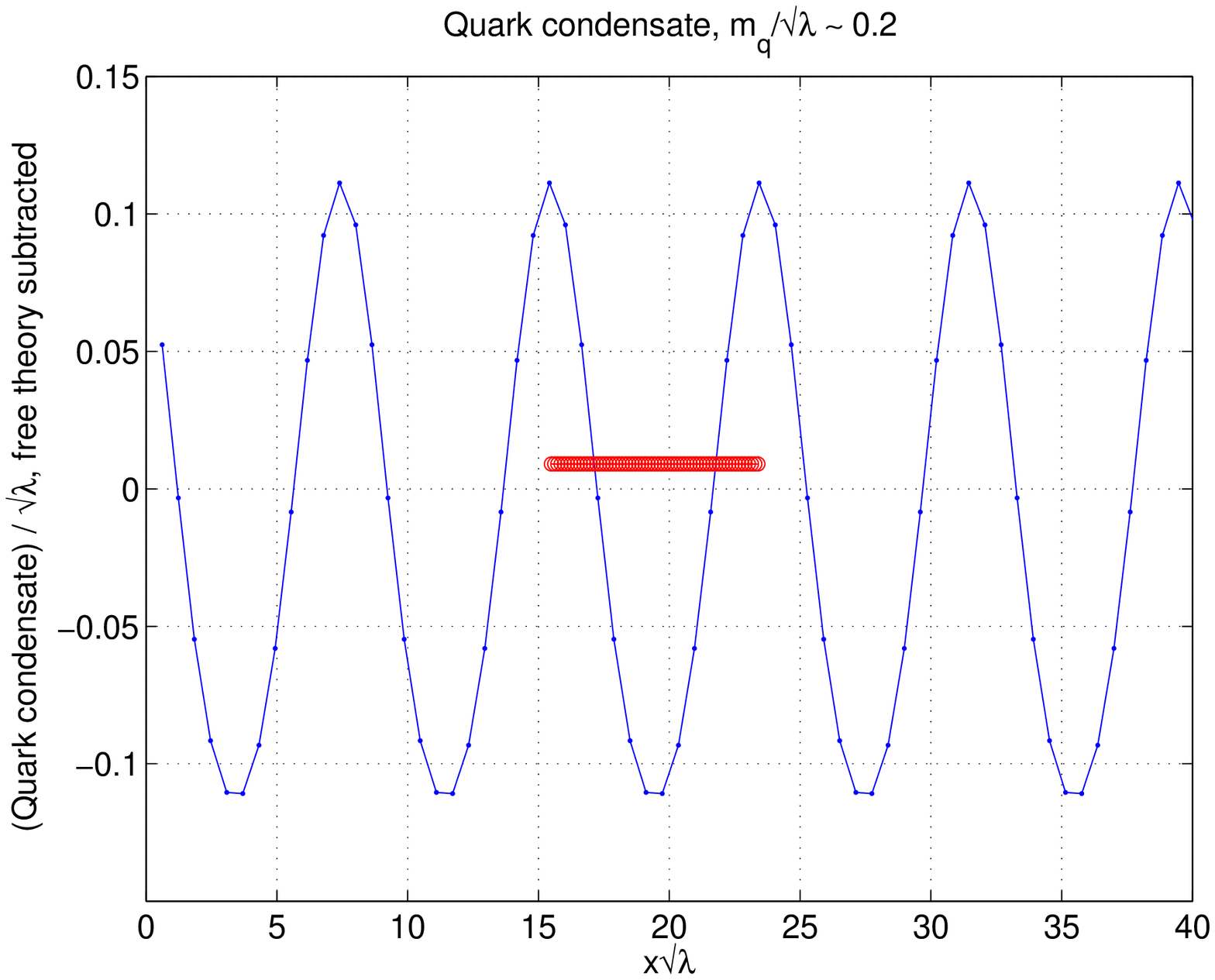}
}
\centerline{
\includegraphics[width=11cm]{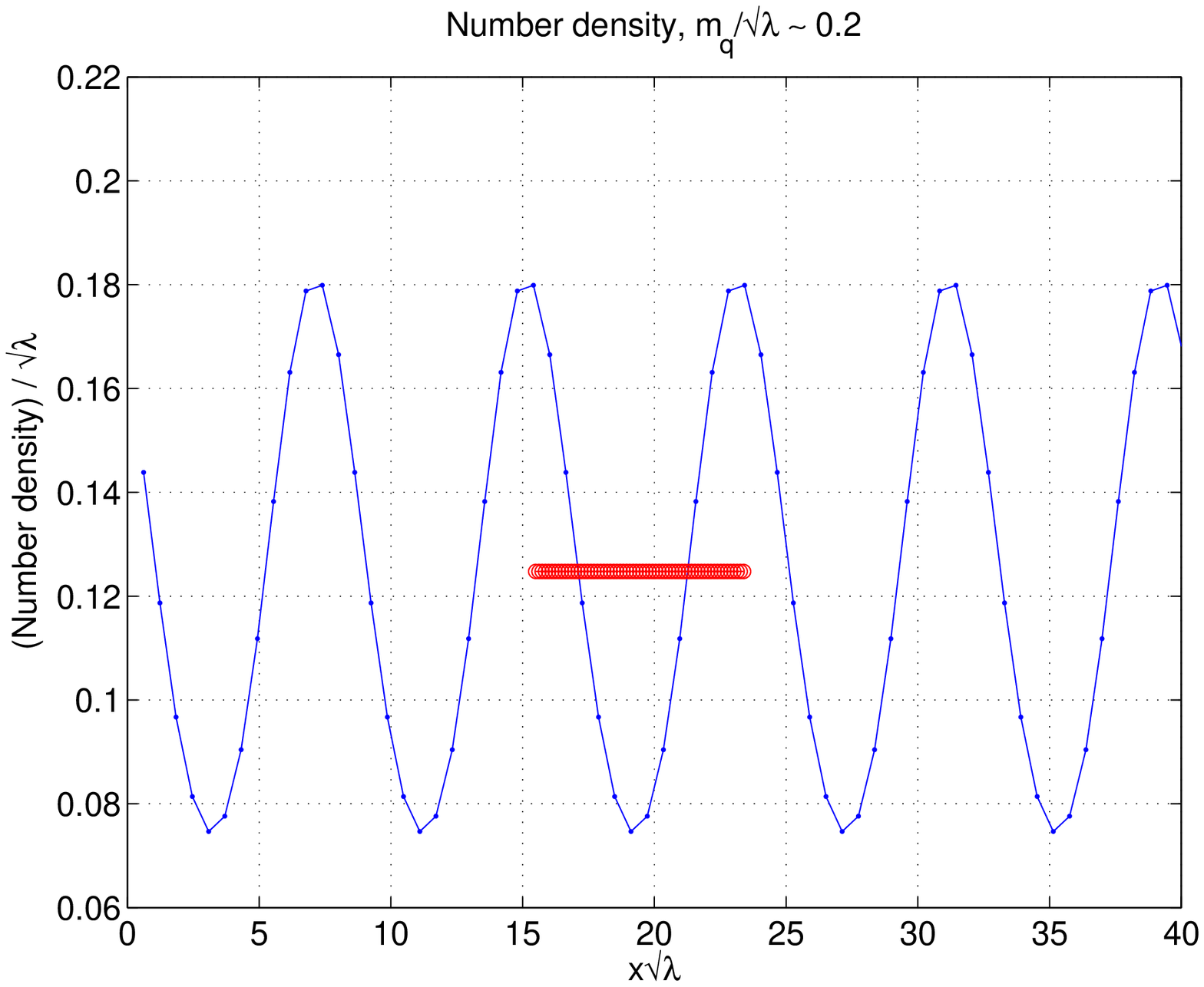}
}
\caption{The way $\<\bar \psi \psi\>$ (top panel) and the baryon density (bottom panel) depend on the spatial coordinate for $M=1$ (i.e. by mistreating the zero modes). Results in dots (blue) are for a  system of $B=5$ in a volume of $L\sqrt{\lambda/(2\pi)}=16$ and results in squares (red) are for $B=1$ and $L'=L/5$. Clearly the mistreatment of the zero modes, reflected by the choice $M=1$, does \underline{not} lead to volume independence -- the squares and the dots do \underline{not} coincide.}
\label{partial1}
\end{figure}

\begin{figure}[p]
\centerline{
\includegraphics[width=12cm]{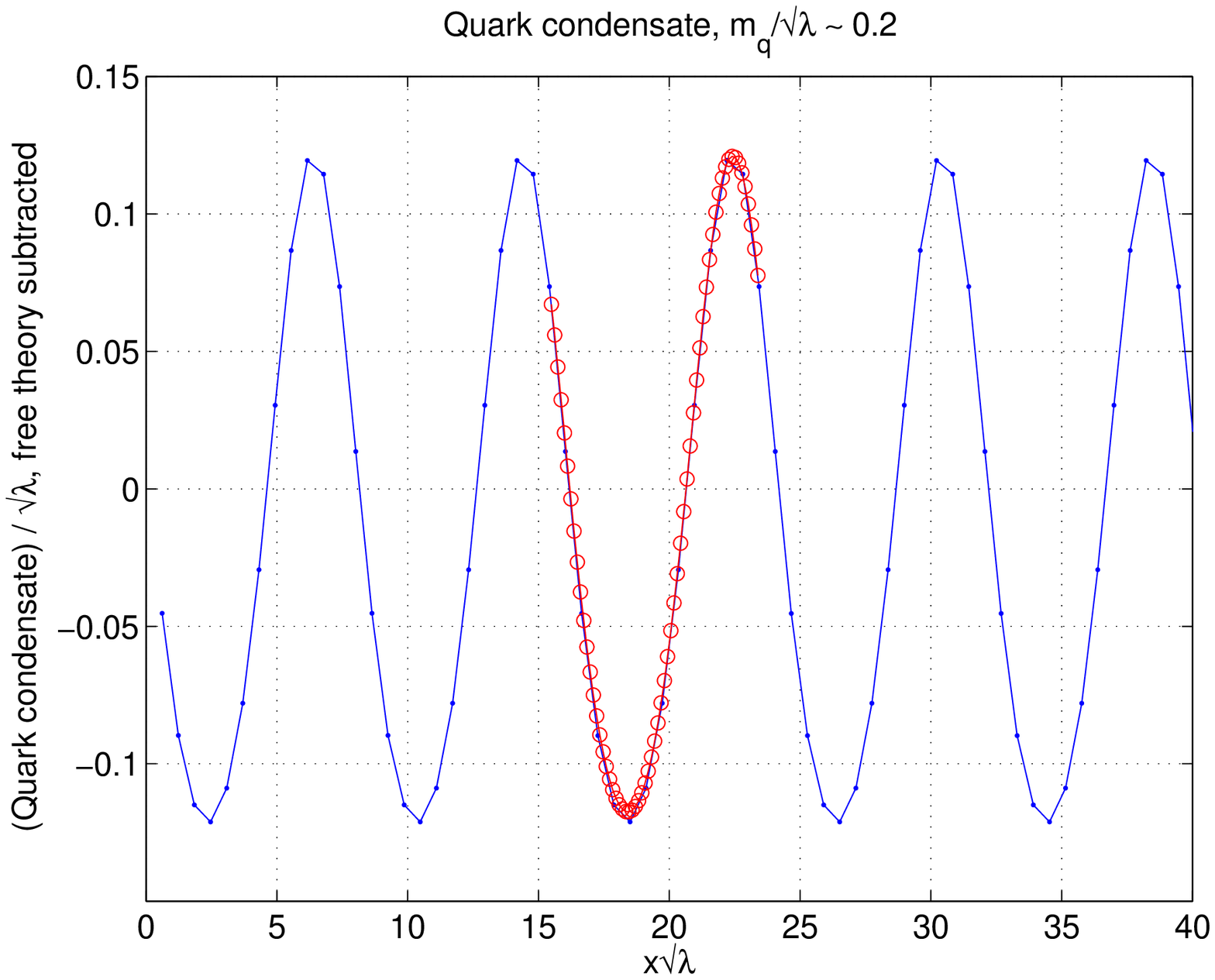}
}
\centerline{
\includegraphics[width=12cm]{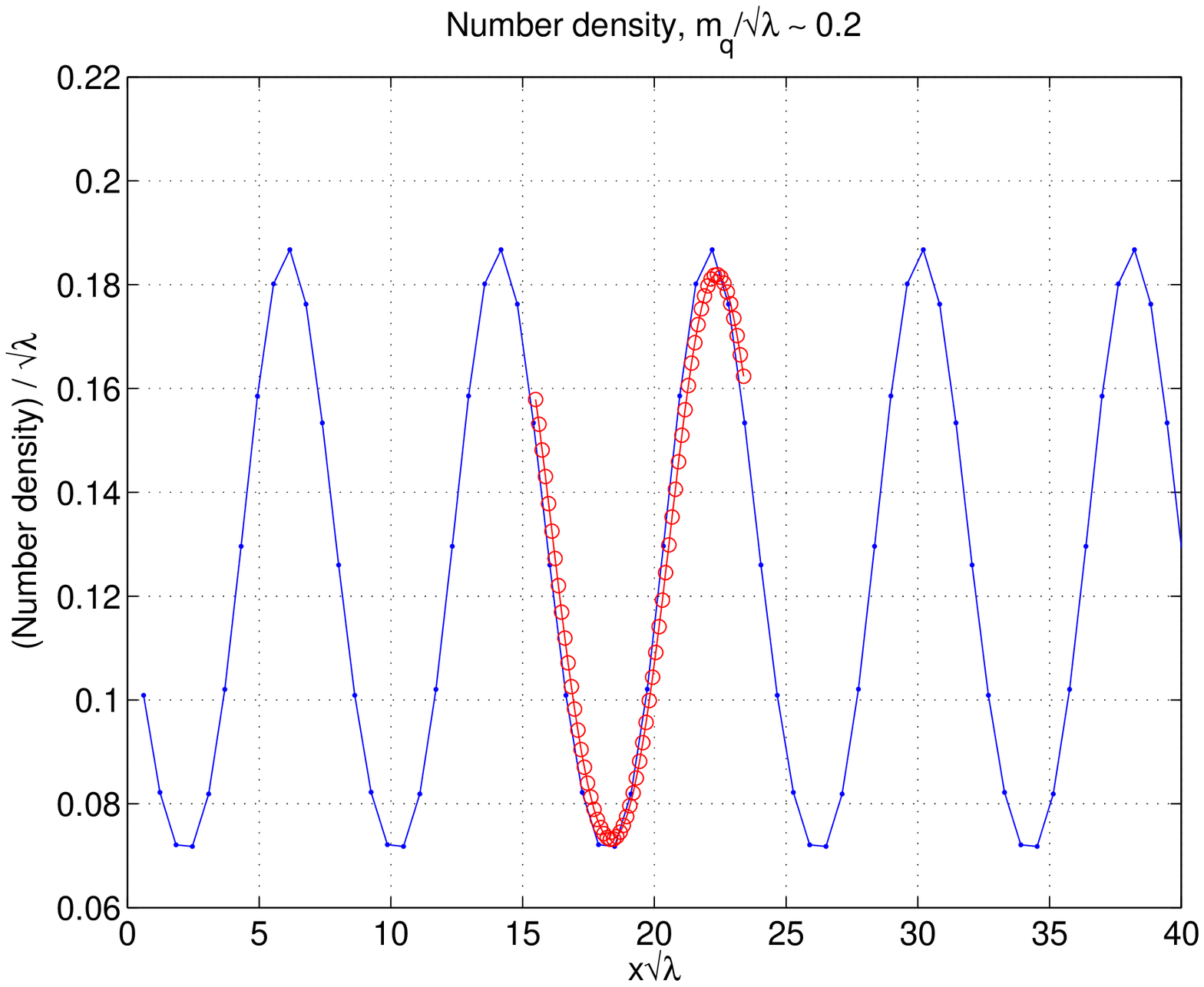}
}
\caption{The same as Fig.~(\ref{partial1}), but when one correctly incorporates the zero modes in the calculation : here $M=25$ rather than $M=1$. In this case the dots and squares coincide and partial large-$N$ volume independence holds.}
\label{partial2}
\end{figure}

The goal of this section is to check the following simple observation that was made in Ref.~\cite{thooftaxial}: Despite the breakdown of translation invariance in the presence of baryons, there is a remnant of the ``Eguchi-Kawai'' volume independence of large-$N$ QCD which allows us to calculate properties of $B$ baryons in a box of size $L$ by putting a single baryon in a box of size $L/B$. To see this `soft' form of large-$N$ volume independence the spatial gluonic zero modes need to be treated correctly, and as explained above this amounts to using large values of $M$. 
To check this observation we compare our numerical solution of a system with $B=5$ baryons and a volume of $L\sqrt{\lambda/(2\pi)}=16$ to the $B=1$ case and $L'=L/5$. We present the results for the quark condensate and density obtained with $M=1$ in Fig.~\ref{partial1} and with $M=25$ in Fig.~\ref{partial2}. As the plots clearly demonstrate such volume independence takes place only for large values of $M$ (i.e. as long as one treats the zero modes correctly). 

Clearly, the observation we make in this section can have an important practical consequence: if it works in four dimensions as well, it can save computational resources for lattice Monte-Carlo simulations. To see this observe that what determines the size of the lattice corrections is the ratio $a/\Delta$, where $a$ is the lattice spacing and $\Delta$ is the crystal wave length. Thus, to put more baryons in a box, we need to decrease the lattice spacing, and fixing the physical volume, this means increasing the number of lattice points, and thus the calculational cost. If however, we assume that the physical baryon is close to the infinite-$N$ baryon, then according to what we find in this section, we can study large densities by putting a single baryon in a box whose physical size is small. This can be accomplished by just decreasing the lattice spacing, but by keeping the number of lattice sites fixed. At any given volume, the ratio $a/\Delta$ is equal to $a/L$, but the latter is equal to $1/N_s$, with $N_s$ denoting the number of sites in one of the lattice directions, which is kept fixed and large.

The usefulness of this proposal is predicated on two assumptions. First we assume that physical, $3$-color, QCD is well approximated by its large-$N$ limit, even in the presence of baryons. We do not see any obvious problems with this assumption. Second, it assumes that in four dimensions the ground state of the dense system is a crystal. Moreover, it assumes that the crystal structure is commensurate with the lattice topology (i.e. that it is a simple cubic). As we say in the previous section, because we are at large-$N$ a crystal is expected to form in any number of dimensions, and at finite-$N$ this crystal may dissolve, but can leave behind a nontrivial structure in correlation functions, that can be studied with the suggestion we make in this section.

The question of whether the formed crystal is simple-cubic is a dynamical one and remains an unproven assumption in our context.

\section{Summary and conclusions}
\label{summary}

In this paper we studied large-$N$ QCD in $1+1$ dimensions and solved for its ground state given arbitrary baryon number $B$, quark mass $m$, and spatial volume $L$. We used the Hamiltonian lattice formalism and regularized the theory in the IR by placing it in a finite spatial box. This IR regularization prevents us from gauging away the spatial gluons completely, and a set of gluonic zero modes remains and plays a crucial role at small volumes and/or large baryon densities. These zero modes were ignored in some studies of this theory at $B=0$ and $B=1$ and here we analyzed the effect such a mistreatment has on the value of quark condensate $\<\bar \psi \psi\>$ at $B=0$, and on the baryon mass. Next we analyzed the ground state of a variety of baryon numbers in the range $0\le B\le 30$, and how it depends on the volume, on the lattice spacing, and on the quark mass. We find that these ground states break translation invariance and form a crystal for large enough densities, and that this crystal is characterized by a helical structure in the quark bilinear condensate densities.
For light quarks our results closely resemble those obtained analytically by Schon and Thies in Ref.~\cite{SchonThies} who used an ansatz that reduces  the `t Hooft equations, for $m/\surd \lambda \ll 1$, to the corresponding equations in the sine-Gordon model. This work also ignores the gluonic zero modes (which seems justifiable since it focused on the infinite volume limit). In any event, such an agreement is not expected a priori because, first, we make no ansatz for the ground state (but rather minimize the classical Hamiltonian of the system numerically), and, second, most of the quark masses we explore are not light, and some are quite heavy with $m/\surd\lambda \simeq 1.2-2.4$. 

\begin{figure}[p]
\centerline{
\includegraphics[width=15cm]{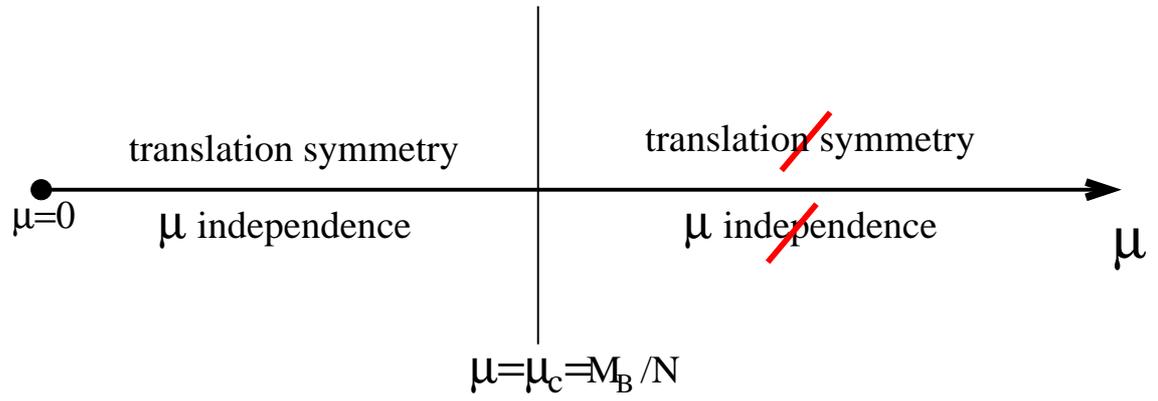}
}
\caption{The phase diagram of two-dimensional large-$N$ QCD in the grand-canonical ensemble. For quark chemical potentials $\mu$ that obey $\mu > m_B/N$, translation symmetry is broken and the ground state depends on $\mu$.}
\label{PD}
\end{figure}

In general, and for any value of $m>0$, our results show that the ground state of large-$N$ QCD in $1+1$ dimensions strongly depends on the baryon number $B$. In particular, for any $B\ge 1$ the ground state breaks translations. To see how this dependence is reflected in the grand-canonical ensemble, we used the way the ground state energies depend on the baryon number and calculated the quark chemical potential $\mu$. This results in the phase structure sketched in Fig.~\ref{PD} : there is a phase transition that occurs at a value of $\mu$ equal to the baryon mass (divided by $N$) and that separates a phase which has zero density, that is $\mu$-independent, and that is invariant to translation symmetry, from a nonzero density phase with spontaneously broken translation symmetry and in which physical observables strongly depend on $\mu$.

Finally, it is interesting that when we calculate the equation of state, we find that for large-$B$, the energy grows quadratically in $B$. This also means that the baryons repel each other, which explains the formation of the crystal. We also see that this quadratic growth is {\em quantitatively} quite close to the quadratic growth in the equation of state of free massless quarks (see Fig.~\ref{dF} and Section~\ref{EoS}). Thus it seems that for a single flavor and in $1+1$ dimensions there is a smooth connection between baryon physics at small/moderate density and free quark physics at large densities.

As mentioned in the introduction, Ref.~\cite{GHN} has attempted to approach large-$N$ two-dimensional QCD by studying, instead of `t Hooft model itself, its associated `zero-volume' Eguchi-Kawai (EK) matrix model \cite{EK}. The main conclusion of Ref.~\cite{GHN} is very puzzling: large-$N$ QCD in $1+1$ dimensions is argued to be completely independent of the baryon chemical potential. This, if correct, contradicts simple physical intuition (and of course contradicts all the results of Refs.~\cite{Salcedo,SchonThies} and of the current paper). Specifically, what the conclusions of Ref.~\cite{GHN} mean is that there are no baryons in two dimensions and so this also contradicts section IX of Ref.~\cite{Witten}. 

One of the reasons why the conclusion of \cite{GHN} is questionable is the following. The use of the EK prescription by the authors of Ref.~\cite{GHN} assumes that the euclidean Dirac operator factorizes in momentum space. This assumption is predicated on having a ground state that is invariant under translations. If the latter symmetry is spontaneously broken in the ground state, then a quark can change its momentum by interacting with momentum-carrying condensates, and such interactions invalidate the factorization property assumed by the authors of Ref.~\cite{GHN}.\footnote{That the Dirac operator does not factorize in momentum space when translations are spontaneously broken is well-known and we refer the reader to \cite{nonzeroQ} for explicit examples.} Put differently, one can easily show that the generating functional of momentum-carrying quark bilinears, such as $\int dx \, \< \bar \psi(x) \, \psi(x) \> \, e^{iQx}$, which break translations, cannot be calculated with the momentum factorization used by Ref.~\cite{GHN}. Indeed, general arguments tell us that Eguchi-Kawai reduction will take place only so long as the ground state is translation invariant (see for example the first paper of Ref.~\cite{YaffeCoherent}), and since we show that for $\mu>m_B/N$ translation symmetry spontaneously breaks, then one cannot use the EK matrix model to study the field theory in that regime.

Another reason why one may question the conclusions of Ref.~\cite{GHN} was exposed in Ref.~\cite{thooftaxial} and is related to general features of the way baryon density would behave within single-site models. The issue is simply stated: if we have any nonzero baryon number $B\ge 1$ on a lattice that has a single site, then, by construction, we force the baryon density to be at the cutoff scale, i.e. of $O(1/{\rm (lattice- spacing)})$. This is true for any lattice coupling, and does not get `better' as we take the naive continuum limit by sending the bare lattice spacing to zero. Thus, by their definition, single-site models can either have zero density, or density at the cutoff scale, and no densities in between. Indeed, this explanation fits well with the findings of Ref.~\cite{GHN}, where a transition from a zero density phase to a nonzero density phase with $O(1/{\rm (lattice- spacing)})$ density is reported. This was also seen in \cite{nonzeroMU}. While Ref.~\cite{GHN} interprets this as having no dependence on the chemical potential in two-dimensional QCD, we suggest that by working on a single site, the approach of Ref.~\cite{GHN} is `forcing' the system to show only this unphysical behavior. It seems reasonable that the real, physical, transition would be seen only if one had a finite number of sites $L_s$, and that to keep lattice artifacts under control, one would have to send $L_s$ to infinity. With such a finite number of sites, one may also be able to see the breakdown of translation symmetry.

\section*{Acknowledgments}
I thank O.~Bergman, A.~Karch, S.~R.~Sharpe, and D.~Son for useful discussions. This work was supported in part by the U.S. Department of Energy under Grant No. DE-FG02-96ER40956.

\appendix

\section{Free theory subtractions}
\label{appfree}

For the purpose of subtracting the UV divergences from expectation values, we calculate the values of the quark condensate for the free theory in this appendix. Our starting point is the differential equation for $\phi$ in the free theory, which we obtain by setting $\hat\lambda=0$ into \Eq{h}. This gives 
\begin{equation}
\frac{i}2 \left(\phi^n_a(x+1) - \phi^n_a(x-1)\right) + \hat m\, (-1)^x\, \phi^n_a(x) = \epsilon^n_a \, \phi^n_a(x). \label{h0}
\end{equation}
It is convenient to define a new lattice with double the lattice spacing and a unit cell that contains two fields. We choose the following convention (from here on we suppress the index $a$ since it enters \Eq{h0} trivially)
\begin{equation}
\phi^n(x) = \left[ 
\begin{array}{cc}
\phi^n_e(X=\frac{x}2) & \qquad x = {\rm even},\\
\phi^n_o(X=\frac{x+1}{2}) & \qquad x = {\rm odd}.
\end{array}
\right.
\end{equation}
In terms of $\phi_{e,o}(X)$, \Eq{h} becomes the following two sets of coupled equations
\begin{eqnarray}
\frac{i}2 \left(\phi^n_e(X) - \phi^n_e(X-1)\right) - \hat m\, \phi^n_o(X) = \epsilon^n_a \, \phi^n_o(X), \label{h0_o}\\
\frac{i}2 \left(\phi^n_o(X+1) - \phi^n_o(X)\right) + \hat m\, \phi^n_e(X) = \epsilon^n_a \, \phi^n_e(X). \label{h0_e}
\end{eqnarray}
Since $X$ obtains the values $1,2,3,\dots,L_s/2$ we Fourier transform it as
\begin{equation}
\phi^n_{o,e}(X) = \frac1{\sqrt{L_s/2}}\, \sum_{p}\, \phi^n_{e,o}(p)\, e^{ipX},
\end{equation}
where here $p=2\pi k /(L_s/2)$ and $k=1,2,3,\dots,L_s/2$. Substituting this into Eqs.~(\ref{h0_o})--(\ref{h0_e}) gives the following matrix equation
\begin{equation}
\left( 
\begin{array}{cc}
-m & -e^{-ip/2}\sin(p/2)\\
-e^{+ip/2}\sin(p/2)& m
\end{array}
\right) \, 
\left(
\begin{array}{c}
\phi^n_o(p)\\
\phi^n_e(p)
\end{array}
\right) = 
\epsilon^n_p\left( 
\begin{array}{c}
\phi^n_o(p)\\
\phi^n_e(p)
\end{array}
\right),
\end{equation}
whose orthonormal eigenvectors are 
\begin{equation}
\left(
\begin{array}{c}
\phi^{\pm}_o(p)\\
\phi^{\pm}_e(p)
\end{array}
\right) = 
\frac1{\sqrt{2\sqrt{\hat m^2 + \sin^2(p/2)}\left(\sqrt{\hat m^2 + \sin^2(p/2)}\pm \hat m\right)}}
\left(
\begin{array}{c}
-e^{-ip/2} \sin(p/2)\\
\hat m \pm\sqrt{\hat m^2 + \sin^2(p/2)} 
\end{array}
\right),\label{eigvec}
\end{equation}
and whose eigenvalues are $\epsilon^\pm_p=\pm \sqrt{\hat m^2 + \sin^2(p/2)}$. 

For a fixed value for the baryon number $B$, the elements of the matrix density $\rho_{xy}$ are set by \Eq{rho} and in our case the index $n$ in that equation corresponds to the combination of the momentum index $p$ and the $\pm$ index of \Eq{eigvec}. Since we are seeking for the solution with minimum energy, the sum over the combined index $n\equiv (p,\pm)$ that forms $\rho_{xy}$ (see \Eq{rho}) includes the following :
\begin{itemize}
\item All the eigenvectors  $\phi^-(p)$ with $p=1,2,\dots,L_s/2$ (all these have negative energy $\epsilon^-(p)$ and so by including all of them we lower the energy to a minimum).
\item A set of $B$ eigenvectors $\phi^+(p)$ whose energy $\epsilon^+(p)$ is the lowest. Below we denote the set of momenta that corresponds to these eigenvectors by  $F$. 
\end{itemize}
Focusing only on the diagonal elements of $\rho$, this results in the following form
\begin{equation}
\rho_{xx} = \frac{1}{L_s/2}\times \left[ \begin{array}{cc}
{\displaystyle \sum_{p}\, |\phi^-_e(p)|^2 + \sum_{p\in F}\, |\phi^+_e(p)|^2} & \qquad x={\rm even},\vspace{0.5cm} \\
{\displaystyle \sum_{p}\, |\phi^-_o(p)|^2 + \sum_{p\in F}\, |\phi^+_o(p)|^2} & \qquad x={\rm odd}.
\end{array}
\right.
\label{rhofree}
\end{equation}
Using Eqs.~(\ref{lat2cont1})--(\ref{lat2cont2}), and after some algebra, we obtain 
\begin{eqnarray}
\<\psi^\dag \psi\>_{\rm free}(X) &=&  \frac1{L_s}\left(\sum_{p}  + \sum_{p\in F}\right) = B/L_s + \frac12,\label{1st} \\
\<\bar \psi \psi\>_{\rm free}(X) &=& \frac{\hat m}{L_s} \left\{ -\sum_p \frac{1}{\sqrt{\hat m^2 + \sin^2 (p/2) }} + \sum_{p\in F} \frac{1}{\sqrt{\hat m^2 + \sin^2 (p/2)}} \right\},\label{2nd}\\
\<\bar \psi \, i\gamma_5 \psi\>_{\rm free}(X) &=& \frac{ 1}{L_s} \left\{ \sum_p \frac{\sin^2(p/2)}{\sqrt{\hat m^2 + \sin^2 (p/2) }} - \sum_{p\in F} \frac{\sin^2(p/2)}{\sqrt{\hat m^2 + \sin^2 (p/2)}} \right\},\label{3rd}\\
\<\bar \psi \gamma_1\psi\>_{\rm free}(X) &=& 
\frac{1}{2L_s} \left\{ \sum_p \frac{\sin(p)}{\sqrt{\hat m^2 + \sin^2 (p/2) }} + \sum_{p\in F} \frac{\sin(p)}{\sqrt{\hat m^2 + \sin^2 (p/2)}} \right\}
.\label{4th}
\end{eqnarray}
\Eq{1st}  tells us to subtract $1/2$ from the density of the interacting theory, hence \Eq{nbfree}. In the infinite volume limit the remainder is
\begin{equation}
\<\psi^\dag \psi(X)\>_{\rm free} =  \frac{1}2 \int_{-p_F}^{p_F} \frac{dp}{2\pi} = B/L_s,
\end{equation}
which gives us the value of the Fermi momentum $p_F=\pi \frac{B}{L_s/2}$. Using this and taking the infinite volume limit of Eqs.~\ref{2nd}--\ref{3rd} we obtain \Eq{condfree} and \Eq{cond5free}. Finally, the symmetry $p\to -p$ of the sums in \Eq{4th} tells us that $\<\bar\psi \, \gamma_1\, \psi\>=0$.

\vfil\eject

\vfil\eject




\end{document}